\documentclass[pdflatex,sn-vancouver-num]{sn-jnl}%

\pdfoutput=1
\usepackage[utf8]{inputenc}
\usepackage[english]{babel}
\usepackage[T1]{fontenc}
\usepackage{amsmath}
\usepackage{amssymb}
\usepackage{amsfonts}
\usepackage{hyperref}
\usepackage{comment}
\usepackage{physics}
\usepackage{booktabs}
\usepackage{mathtools}
\usepackage{lipsum}
\usepackage[normalem]{ulem}
\usepackage{xcolor}

\theoremstyle{thmstyleone}%

\theoremstyle{thmstyletwo}%

\theoremstyle{thmstylethree}%

\newif\iftodoon
\todoontrue %

\newcommand{\projector}[2]{\ket{#1}\bra{#1}_{#2}}
\newcommand{\solutionString}{\mathbf{s}}
\newcommand{\optimalSolution}{\solutionString^{*}}
\newcommand{\identity}{\mathbb{I}}
\newcommand{\projectorExpanded}[2]{\dfrac{\identity_{#2} + (-1)^{#1} \hat{Z}_{#2}}{2}}
\newcommand{\autocite}[1]{\cite{#1}}

\newcommand{\figureLabel}{Fig.~}
\newcommand{\sectionLabel}{Sec.~}
\newcommand{\equationLabel}{Eq.~}

\newcommand{\response}[1]{#1}

\begin{document}

\title[Article Title]{Resource-Efficient Quantum Optimization via Higher-Order Encoding}

\author*[1]{\fnm{Frederik} \sur{Koch}}\email{frederik.koch@uni-hamburg.de}

\author[1,2]{\fnm{Shahram} \sur{Panahiyan}}\email{shahram.panahiyan@uni-hamburg.de}

\author[1,3,4,5]{\fnm{Rick} \sur{Mukherjee}}\email{rick-mukherjee@utc.edu}

\author[6]{\fnm{Joseph} \sur{Doetsch}}\email{joseph.doetsch@lhind.dlh.de}

\author[1,7,8]{\fnm{Dieter} \sur{Jaksch}}\email{dieter.jaksch@uni-hamburg.de}

\affil*[1]{\orgdiv{Institute for Quantum Physics}, \orgname{University of Hamburg}, \orgaddress{\street{Luruper Chaussee 149}, \city{Hamburg}, \postcode{22761}, \state{Hamburg}, \country{Germany}}}

\affil[2]{\orgdiv{Max Planck Institute for the Structure and Dynamics of Matter}, \orgaddress{\street{Luruper Chaussee 149}, \city{Hamburg}, \postcode{22761}, \state{Hamburg}, \country{Germany}}}

\affil[3]{\orgdiv{Zentrum für Optische Quantentechnologien}, \orgname{University of Hamburg}, \orgaddress{\street{Luruper Chaussee 149}, \city{Hamburg}, \postcode{22761}, \state{Hamburg}, \country{Germany}}}

\affil[4]{\orgdiv{Department of Physics \& Astronomy}, \orgname{University of Tennessee}, \orgaddress{\street{TN 37403}, \city{Chattanooga}, \postcode{37403}, \state{Tennessee}, \country{USA}}}

\affil[5]{\orgdiv{UTC Quantum Center}, \orgname{University of Tennessee}, \orgaddress{\street{TN 37403}, \city{Chattanooga}, \postcode{37403}, \state{Tennessee}, \country{USA}}}

\affil[6]{\orgdiv{Lufthansa Industry Solutions}, \orgaddress{\street{Südportal 7}, \city{Norderstedt}, \postcode{22848}, \state{Schleswig-Holstein}, \country{Germany}}}

\affil[7]{\orgdiv{The Hamburg Centre for Ultrafast Imaging}, \orgname{University of Hamburg}, \orgaddress{\street{Luruper Chaussee 149}, \city{Hamburg}, \postcode{22761}, \state{Hamburg}, \country{Germany}}}

\affil[8]{\orgdiv{Clarendon Laboratory}, \orgname{University of Oxford}, \orgaddress{\street{Parks Road}, \city{Oxford}, \postcode{OX1 3PU}, \state{Oxford}, \country{United Kingdom}}}

\abstract{Quantum approaches to combinatorial optimization problems (COPs) are often limited by the resource demands of Quadratic Unconstrained Binary Optimization (QUBO) encodings, which enlarge circuits through penalty terms and increase qubit and gate counts. We show that Higher-Order Unconstrained Binary Optimization (HUBO) enables a more resource-efficient formulation. Our method systematically constructs HUBO Hamiltonians and, compared to \response{a QUBO formulation} in benchmarks on Gate Assignment (GAP), Maximum k-Colorable Subgraph (MkCS), and Integer Programming (IP) problems, \response{significantly} reduces qubit requirements and decreases \response{total} CNOT gate counts by at least 89.6\% for all tested instances. These results highlight HUBO as a practical alternative for quantum optimization on near-term devices. To promote adoption, we release an open‑source Python library that automates HUBO model construction, extends beyond the examples presented in this work, and broadens access to resource‑efficient quantum optimization.}


\maketitle

\section{Introduction}

Combinatorial optimization problems are prominent in numerous scientific and industrial domains, ranging from logistics and finance \autocite{marzecPortfolioOptimizationApplications2016} to computational biology \autocite{perdomo-ortizFindingLowenergyConformations2012} and operations research \autocite{doi:10.1142/12343}. Challenging problems require finding (near)-optimal solutions within vast discrete configuration spaces, where the search complexity often grows exponentially with problem size \autocite{crescenziCompendiumNPOptimization1995}. Theoretical insights from statistical physics have illuminated these challenges through their connection to disordered systems, revealing that most industrial optimization problems belong to the NP-hard complexity class \autocite{fuApplicationStatisticalMechanics1986}.

Classical methods, including metaheuristics such as simulated annealing and tabu search \cite{kirkpatrickOptimizationSimulatedAnnealing,gloverUsersGuideTabu1993}, and commercial solvers like CPLEX and Gurobi \cite{IBMILOGCPLEX2024a, LeaderDecisionIntelligence} face significant limitations when confronting non-convex energy landscapes. Even relatively small problems (sometimes below 100 variables) remain computationally intractable for them \autocite{xuRelaxationsBinaryPolynomial2024, puchingerMultidimensionalKnapsackProblem2010,packebuschLowAutocorrelationBinary2016, danilovaRecentTheoreticalAdvances2022, burerNonconvexMixedintegerNonlinear2012,floudasGlobalOptimization21st2005}.

Quantum algorithms, such as adiabatic quantum optimization \autocite{albashAdiabaticQuantumComputation2018a,ebadiQuantumOptimizationMaximum2022} and the Quantum Approximate Optimization Algorithm (QAOA) \cite{farhiQuantumApproximateOptimization2014}, \response{a digitized, trotterized variant of adiabatic quantum computing,} are poised to address these challenges. This is predominantly done by exploiting mappings of COPs to QUBO models \autocite{lucasIsingFormulationsMany2014, goswamiSolvingOptimizationProblems2024, lai2024arbitraryqubooptimizationanalysis} which is further facilitated by modeling tools such as Pyqubo \cite{zaman2021pyqubo}.
Previous studies \cite{Dominguez_2023, montanez-barreraUnbalancedPenalizationNew2024} have identified several practical limitations of QUBO formulations, including large qubit requirements, costly penalty terms for constraint enforcement, and consequently increased circuit depth with large CNOT gates, which motivate the search for more resource-efficient alternatives.

In contrast, \response{binary} encodings, \response{that give rise to HUBO Hamiltonians,} directly incorporate such constraints and, therefore, can reduce resource demands. \response{At an intuitive level, each decision variable is represented by a compact binary register, so the single-value assignment is native to the encoding rather than enforced by an additional one-hot penalty term.} While demonstrated in specific settings, including protein folding and the traveling salesperson problem \cite{romeroBiasFieldDigitizedCounterdiabatic2024,glosSpaceefficientBinaryOptimization2022,romeroProteinFoldingAlltoall2025, chai2023simulatingflightgateassignment, nagiesBoostingQuantumAnnealing2025}, prior work has lacked a general framework. Moreover, practical concerns remain regarding hardware implementation: higher-order terms formally require multiqubit interactions, whereas present quantum processors primarily support operations involving just one or two qubits at a time \cite{Wintersperger_2023, fausewehQuantumManybodySimulations2024}.
\response{In this work, we use standard decompositions so that such circuits are compiled entirely into single- and two-qubit gates.}

\begin{figure}[t]
    \centering
    \includegraphics[width=85mm]{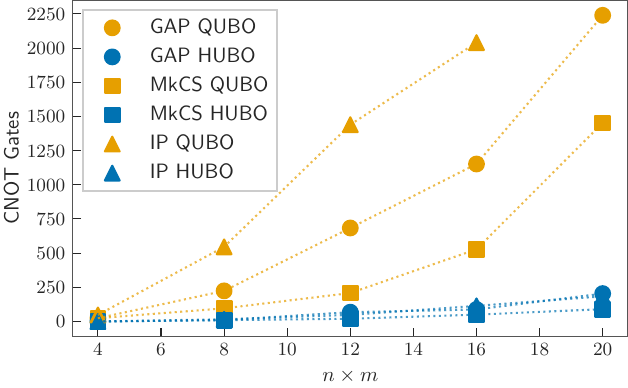}
    \caption{Number of CNOT gates required to reach solution threshold for three COPs, optimized by QAOA with QUBO (orange marker) and HUBO (blue marker) \response{formulations} as a function of problem size $n\times m$ \response{, with $m$ held fixed}. The GAP optimizes the assignment of $n$ aircraft to $m$ airport gates. In the MkCS problem, $n$ vertices are colored with $m$ colors, IP optimizes a polynomial over $n$ variables, each taking one of $m$ possible values. Further details and quantum encodings are provided in the main text.
    }
    \label{fig:first_figure_cnot_gates}
\end{figure}

By comparing to \response{a} QUBO \response{formulation}, we demonstrate that \response{the HUBO formulations presented in this work} provide substantial and consistent resource advantages for three representative COPs: GAP, MkCS, and IP. \response{These advantages persist after full compilation to native one- and two-qubit gates, with HUBO significantly lowering qubit counts and reducing CNOT requirements by at least 89.6\% for all benchmarked problem classes and sizes (see Figure~\ref{fig:first_figure_cnot_gates}), i.e., without relying on hardware-level higher-order gates.}
To support broad adoption, we present PyHUBO, an open-source Python library \cite{Koch_PyHUBO_2025}, and introduce a systematic framework for constructing HUBO Hamiltonians \response{through binary encodings}, showing that for the problems considered, our method yields numerical complexity that scales polynomially with problem size.
Although we focus on QAOA, the approach applies to other algorithms and problems with one-hot or related constraints. Our results demonstrate that HUBO \response{formulations} can directly mitigate resource bottlenecks in quantum combinatorial optimization.

\section{Methodology}
\label{sec:Methodology}

The methodology of this work is structured as follows. In \sectionLabel (\ref{sec:formal_definition}), we establish a formal definition and notation for COPs and illustrate this formulation using three representative examples: the GAP, the MkCS, and IP. In \sectionLabel (\ref{sec:quantum_encoding}), we show how the COP formulations from \sectionLabel (\ref{sec:formal_definition}) can be systematically mapped to QUBO and HUBO \response{Hamiltonians}. Finally, \sectionLabel (\ref{sec:quantum_implementation}) describes the implementation of these \response{Hamiltonians} within the QAOA algorithm, utilizing only single- and two-qubit gates.
\subsection{Combinatorial Optimization Problems}
\label{sec:formal_definition}
A COP consists of a finite set of feasible solutions and an objective function to be optimized \cite{schrijverCombinatorailOptimization}.
For the remainder of this work, we use the following definition:
Consider an $n$ variable COP, where every variable can assume one of $m$ values. Variables are indexed by $1 \leq i \leq n$ and values by $1 \leq v \leq m$.
In the following, we will use the terms $i$th variable and variable $i$ interchangeably to refer to the variable indexed by $i$, and similarly, we will use the $v$th value and value $v$ interchangeably to refer to the value indexed by $v$.
A solution $\solutionString \in \{1, \cdots, m\}^{n}$ assigns a value to each variable. For every $\solutionString$ there exists an objective function $C(\solutionString)$ that can be computed in polynomial time and a polynomial time oracle that verifies whether $\solutionString$ is a feasible solution.

The task is to find a solution such that the objective function is minimal among all feasible solutions, i.e., $\optimalSolution = \operatorname*{argmin}_{\solutionString} C(\solutionString)$.
In most applications, the objective function is a polynomial function of the variables \cite{doi:10.1142/12343}, and for practical purposes, we express it as
\begin{equation}
    C(\solutionString) = \sum_{1 \leq i \leq n} c_1(i, s_i) + \sum_{1\leq i<j\leq n} c_2(i, j, s_i, s_j)+ \cdots,
    \label{eqn:classical_objective_function}
\end{equation}
where $c_1(i, s_i)$ is the cost of assigning the value $s_i$ to the $i$th variable, $c_2(i, j, s_i, s_j)$ the cost of assigning the value $s_i$ to the $i$th variable while simultaneously assigning the value $s_j$ to the $j$th variable.

In principle, the objective function can be of any order. However, for most industrial applications, it is linear or quadratic \autocite{crainicCombinatorialOptimizationApplications2024}, i.e. Eq. \eqref{eqn:classical_objective_function} only contains terms of the form $c_1(i, s_i)$ or $c_2(i, j, s_i, s_j)$. \response{Genuinely higher-order terms (required for example in \cite{salehi_unconstrained_2022, lopez-ibanez_travelling_2013}) can be incorporated in the same framework; here we restrict to linear and quadratic forms to keep the presentation concise and the scaling analysis transparent.}

Often, COPs also involve constraints. For practical purposes, we formulate these constraints as inequalities, which can be converted to equality constraints, e.g., using slack variables \cite{chenSlackvariableApproachVariational2025}. The equality constraints are added as penalty terms to \equationLabel \eqref{eqn:classical_objective_function} with appropriately chosen Lagrange multipliers \cite{gloverQuantumBridgeAnalytics2022}.

For example, a COP with a constraint that all variable pairs $(i,j)\in P$ for some set $P$ can not assume the same value can be stated as
\begin{equation}
    \begin{split}
         & \underset{\solutionString \in \{1, \cdots, m\}^{n}}{\text{minimize}}   \quad C(\solutionString) \\
         & \underset{\forall (i,j) \in P}{\text{subject to}} \quad s_i \neq s_j.
    \end{split}
\end{equation}
The penalty term for this constraint has the form
\begin{equation}
    \begin{split}
        c_2(i,j, v, v) & = \lambda \\ \forall (i,j) \in P \quad \forall v &\in \{1, \ldots, m\},
    \end{split}
    \label{eqn:nonequal_constraint}
\end{equation}
with the Lagrange multiplier $\lambda$ chosen large enough such that the optimal solution respects the constraint. By adding this penalty term to the objective function, we can transform the COP into an unconstrained optimization problem
\begin{equation}
    \underset{\solutionString \in \{1, \cdots, m\}^{n}}{\text{minimize}}   \quad C(\solutionString) + \lambda \sum_{(i,j) \in P} \delta_{s_i, s_j},
\end{equation}
where $\delta_{s_i, s_j}$ is the Kronecker delta function, which is 1 if $s_i = s_j$ and 0 otherwise.

The above notation applies broadly to COPs. We now illustrate this on three representative classes: the GAP, the MkCS problem, and IP. These formulations provide the basis for deriving their QUBO and HUBO encodings.

\subsubsection{Gate Assignment Problem}
\label{sec:gate_assignment_problem}
\begin{figure}
    \centering
    \includegraphics[width=85mm]{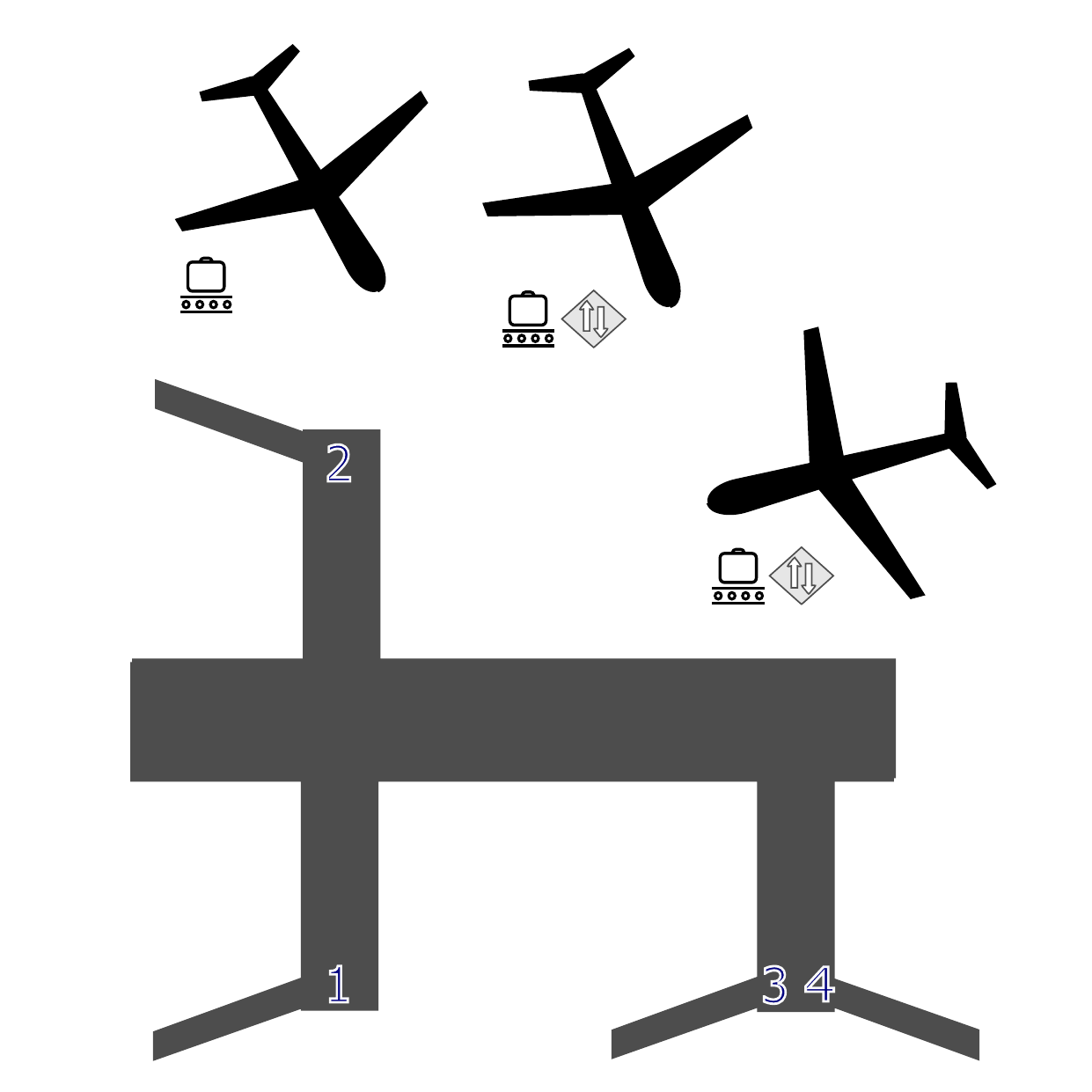}
    \caption{Illustration of the GAP with $m=4$ airport-gates and $n=3$ flights. Two of the flights have passengers who require a transfer connection, which is represented by the two-way symbol. A third flight has only arriving and departing passengers, illustrated by the luggage symbol. The goal is to assign airport-gates to flight such that the total passenger walking time is minimized.}
    \label{fig:airport_illustration}
\end{figure}
The GAP is a well-studied challenge in operations research with significant practical relevance to the aviation industry \cite{bourasAirportGateAssignment2014}. It involves optimally assigning airport-gates to aircraft to minimize costs, such as passenger walking distances.

Various classical formulations of the GAP are established \cite{bourasAirportGateAssignment2014}, and the problem has been studied in the context of quantum optimization \cite{chaiFindingOptimalFlight2023, chai2023simulatingflightgateassignment}. We consider the GAP with $n$ flights and $m$ airport-gates, where the objective is to minimize total passenger walking times (see \figureLabel \ref{fig:airport_illustration}).

We distinguish between two passenger types: those boarding or disembarking, and those transferring between flights.
The first type only contributes to the linear coefficients, as the cost of assigning flight $i$ to airport-gate $v$ is given by
\begin{equation}
    c_1(i, v) = p_i^{\text{arr}} t_{v}^{\text{arr}} + p_i^{\text{dep}} t_{v}^{\text{dep}},
    \label{eqn:gap_linear_term}
\end{equation}
where $p_i^{\text{arr}}$ and $p_i^{\text{dep}}$ are the numbers of arriving and departing passengers on flight $i$, and $t_{v}^{\text{arr}}$ and $t_{v}^{\text{dep}}$ are the walking times from the check-in desk to the airport-gate $v$ and from the airport-gate $v$ to the luggage claim, respectively.

The transfer passengers contribute to the quadratic terms through
\begin{equation}
    c_2(i, j, v, w) = p_{i,j}^{\text{trans}} t_{v, w}^{\text{trans}},
    \label{eqn:gap_quadratic_term}
\end{equation}
where $p_{i,j}^{\text{trans}}$ is the number of passengers transferring from flight $i$ to flight $j$, and $t_{v, w}^{\text{trans}}$ is the walking time between airport-gates $v$ and $w$.

Finally, we enforce the constraint that temporally overlapping flights cannot share the same airport-gate through a penalty term of the same form as in \equationLabel \eqref{eqn:nonequal_constraint}, where $P$ in the context of the GAP is the set of all conflicting flight pairs.

\subsubsection{Maximum k-Colorable Subgraph Problem}
\label{sec:graph_coloring}
\begin{figure}
    \centering
    \includegraphics[width=85mm]{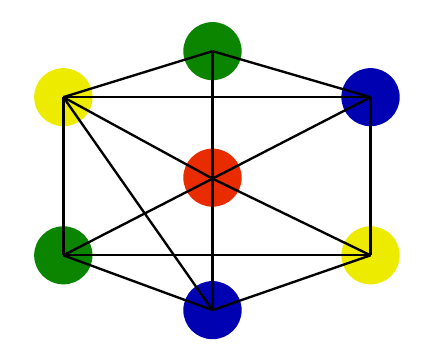}
    \caption{Example instance for the MkCS problem: a properly 4-colored graph with $|V|=7$. All adjacent vertices have distinct colors, so the objective value (number of monochromatic edges) is zero.}
    \label{fig:ExampleMKCS}
\end{figure}
The MkCS \response{problem} represents a fundamental COP that arises in diverse industrial applications, including network design, frequency assignment, and resource allocation \cite{bentertInductive$k$independentGraphs2019, halldorssonSpectrumSharingGames2004, hertzConstructiveAlgorithmsPartial2016, RoutingNetworkDimensioning2006}. It has attracted considerable attention in both classical and quantum optimization research \cite{liuEfficientHybridVariational2025, quinteroCharacterizationQUBOReformulations2021, wang$XY$mixersAnalyticalNumerical2020, streifQuantumAlgorithmsLocal2021, sotirovMaximum$k$colorableSubgraph2021}.

Given a graph $G = (V, E)$ with vertices $V$ and edges $E$, the MkCS problem seeks to maximize the size of a properly vertex-colored subgraph with at most $k$ colors (see \figureLabel \ref{fig:ExampleMKCS}). A properly vertex-colored subgraph is one where no two adjacent vertices share the same color, and the size of the subgraph is defined as the number of edges in the subgraph.

In the assignment formulation, each vertex is a variable that can assume $k$ values, hence $n = \abs{V}$ and $m=k$. The solution string $\solutionString$ encodes the color of each vertex and $s_i$ is the index for the color assigned to the vertex with index $i$. The objective function counts the number of edges in the graph that connect vertices of the same color. Minimizing the number of monochromatic edges is equivalent to maximizing the number of edges that connect vertices of different colors, which is the goal of the MkCS problem.
Using the notation from \sectionLabel \eqref{sec:formal_definition}, the objective function has the coefficients
\begin{equation}
    c_2(i, j, v, v) = 1 \quad  \forall (i, j) \in E,
\end{equation}
and all other coefficients are zero.

\subsubsection{Integer Programming}
\label{sec:ip}
IP is fundamental to industrial applications such as production planning and resource allocation, as well as a key subject in classical optimization research \cite{Formulations2020, ProductionPlanningMixed2006, magataoMixedIntegerProgramming2002}. Recently, quantum and hybrid methods have been employed to address IP problems \cite{goswamiSolvingOptimizationProblems2024, goswamiQuditbasedScalableQuantum2025, svenssonHybridQuantumClassicalHeuristic2023a}.

Here, we consider an IP problem with $n$ variables and each variable can assume any value from the domain $\{y_1, \cdots, y_m \}$ with $y_i \in \mathbb{Z}$. The task is to assign values to variables such that
\begin{equation}
    \underset{\mathbf{u} \in \{y_1, \dots, y_m\}^n}{\text{minimize}} \quad \mathbf{q}^T \mathbf{u} + \mathbf{u}^T Q \mathbf{u} + \cdots,
    \label{eqn:ip_objective_function}
\end{equation}
where $\mathbf{q} \in \mathbb{R}^n$ and $Q \in \mathbb{R}^{n \times n}$ are the coefficients of the objective function. In principle, the objective function \equationLabel \eqref{eqn:ip_objective_function} can be of any order. However, for the sake of clarity, here we write out only up to quadratic order. Typically, an IP problem also involves inequality constraints, which often are functions of many variables.

The objective function can be put into the assignment formulation of Sec. (\ref{sec:formal_definition}) by writing
\begin{equation}
    \begin{split}
        c_1(i, v)       & = q_i y_v,         \\
        c_2(i, j, v, w) & = Q_{i j} y_v y_w,
    \end{split}
    \label{eqn:ip_objective_assignment_coeff}
\end{equation}
for all $1 \leq v, w \leq m$. We find that computing the coefficients of the objective functions under consideration has computational complexity $\mathcal{O}\left(\binom{n}{2} m^{2}\right)$ (since in general there are $\binom{n}{2} m^{2}$ quadratic and $n\times m$ linear coefficients).

A range of methods exists to enforce the constraints of IP problems. Penalty-based techniques are most common, encoding constraints as additional terms in the objective function. Examples include slack variable methods that introduce additional variables to convert inequalities into penalizable equalities ~\cite{chenSlackvariableApproachVariational2025}, as well as more recent variants such as unbalanced penalization~\cite{montanez-barreraUnbalancedPenalizationNew2024}, and Lagrangian approaches ~\cite{sharmaCuttingSlackQuantum2025}, both of which avoid increasing the variable space. Alternatively, oracle-based methods verify feasibility directly, without penalties~\cite{goswamiQuditbasedScalableQuantum2025}. Together, these strategies are broadly applicable to general constraint enforcement in IP.

In our benchmark examples, we restrict ourselves to constraints that each involve at most two variables. For such simple pairwise constraints, it is most straightforward to explicitly enumerate all violating value pairs, which can be done efficiently with computational complexity $\mathcal{O}(m^2)$ per constraint. Once identified, for each pair of values $y_v, y_w$ violating a constraint with variables $i$ and $j$ the objective function coefficients are changed
\begin{equation}
    c_2(i,j,v, w) \rightarrow c_2(i,j,v, w) + \lambda,
\end{equation}
with a sufficiently large Lagrange multiplier $\lambda$. This direct penalty-based formulation is particularly efficient and effective for enforcing simple pairwise constraints in the  framework considered here.

\subsection{From Classical to Quantum Optimization}
\label{sec:quantum_encoding}
The task of mapping a COP to its quantum formulation is twofold. Firstly, we construct a unique mapping between classical solutions and quantum states $\solutionString \leftrightarrow \ket{\mathbf{x}}$. Secondly, we construct a Hamiltonian $\hat{H}_C$ such that $\hat{H}_C\ket{\mathbf{x}} = C(\mathbf{x}) \ket{\mathbf{x}}$. We define $C(\mathbf{x}) \leftrightarrow C(\solutionString)$ in which $C(\solutionString)$ is given by \equationLabel \eqref{eqn:classical_objective_function}. Finding the optimal solution to the COP corresponds to finding the ground state of the Hamiltonian.
In the following, we discuss two different approaches to this mapping \response{yielding QUBO and HUBO Hamiltonians respectively}: the \response{One-Hot} and the \response{Binary encoding}.
\subsubsection{One-Hot Encoding/QUBO Hamiltonian}
\label{sec:one_hot_encoding}
In the \response{One-Hot encoding}, for each variable value pair $(i,v)$ a binary variable $x_{i, v}$ is created.
The $i$th variable taking the $v$th value, is encoded by setting the binary variable $x_{i,v} = 1$, while all other binary variables $x_{i,w}$ with $w \neq v$ are set to $0$.

The objective function \equationLabel \eqref{eqn:classical_objective_function} can be written in terms of binary variables $x_{i, v}$ as
\begin{equation}
    \begin{split}
        C(\mathbf{x}) & = \sum_{1\leq i \leq n} \sum_{1\leq v \leq m} c_1(i, v) x_{i, v} \\ &+ \sum_{1 \leq i<j\leq n}  \sum_{1 \leq v, w \leq m} c_2(i,j, v, w) x_{i,v} x_{j,w}.
    \end{split}
    \label{eqn:one_hot_objective_function}
\end{equation}

To ensure that each variable $i$ can only take one value, we must add a so-called one-hot penalty term
\begin{equation}
    \lambda \left(1 - \sum_{v=1}^{m}  x_{i, v} \right)^2,
    \label{eqn:one_hot_constraint}
\end{equation}
to the objective function, which ensures that for each variable $i$ exactly one of the binary variables $x_{i, v}$ is 1 and all others are 0.
For constrained COPs, other penalty terms need to be added to the cost function to ensure that the constraints are satisfied. For example, the constraint introduced in Sec.~(\ref{sec:formal_definition}), %
which requires that all variable pairs $(i,j)\in P$ for some set $P$ cannot assume the same value, is enforced by the penalty term
\begin{equation}
    \lambda \sum_{i,j \in P} \sum_{v=1}^{m} x_{i, v} x_{j, v}.
\end{equation}
Similarly, other constraints can be added as penalty terms to the objective function. A discussion about the construction of the penalty terms can be found in \autocite{gloverQuantumBridgeAnalytics2022}.

From the \response{One-Hot formulation}, a quantum system is constructed where each One-Hot variable is mapped to one qubit. Then, the value of the qubit \response{$\ket{b}_{i,v}$ with $b\in\{0,1\}$} in the computational basis represents the value of the One-Hot variable $x_{i,v}$. \response{Since there are $n$ variables and $m$ values per variable, the One-Hot encoding uses exactly $n \times m$ qubits (one qubit for each pair $(i,v)$).} Now the Hamiltonian for this encoding can be found by substituting $x_{i, v} \rightarrow (\identity - \hat{Z}_{i, v})/2$, where $\hat{Z}_{i,v}$ is the Pauli-Z operator acting on the qubit encoding the variable $x_{i, v}$. This gives the operator form of the \response{QUBO} Hamiltonian
\begin{equation}
    \begin{split}
        \hat{H}_C & = \sum_{1 \leq i \leq n}\sum_{1 \leq v \leq m} J_{i,v} \hat{Z}_{i, v} \\&+ \sum_{1 \leq i\response{,}j \leq n}\sum_{1 \leq v,w \leq m} J_{i,j,v,w} \hat{Z}_{i, v} \hat{Z}_{j, w},
    \end{split}
    \label{eqn:one_hot_hamiltonian}
\end{equation}
where the linear cost coefficients $c_1(i, v)$ contribute to the linear QUBO coefficients $J_{i,v}$ and the quadratic cost coefficients $c_2(i,j, v, w)$ contribute to $J_{i,v}$, $J_{j,w}$ and $J_{i,j,v,w}$. It is easy to check that with this Hamiltonian we have $\hat{H}_C\ket{\mathbf{x}} = C(\mathbf{x})\ket{\mathbf{x}}$.
In most numerical studies, the QUBO coefficients $J_{i,v}$ and $J_{i,j,v,w}$ are calculated from $c_1(i, v)$ and $c_2(i,j, v, w)$ through packages such as pyqubo \cite{zaman2021pyqubo, tanahashi2019application}.

\response{The One-Hot encoding used here is the most widely adopted encoding that yields a QUBO Hamiltonian. We note, however, that this correspondence is not unique: other encodings (for example, Domain-Wall encoding \cite{Dominguez_2023}) can also produce QUBO Hamiltonians. In this work, we focus on One-Hot because of its broad use in the literature. Within this scope we use the terms One-Hot encoding and QUBO formulation interchangeably.}

\subsubsection{Binary Encoding/HUBO Hamiltonian}
\label{sec:higher_order_encoding}
In the Binary encoding, each variable $i$ is represented using $d = \lceil \log_2(m) \rceil$ bits. To retrieve the value assigned to variable $i$, we read its associated $d$ bits, convert the binary string to the decimal value $v$, and conclude that variable $i$ assumes the value \response{$v+1$ (The $+1$ is introduced so the bitstring $\vec{0}$ encodes the value with index $1$).}

For a quantum system with $n\times d$ qubits encoding the COP solution, measuring the $d$-qubit state \response{$\ket{\mathbf{b}}_i \equiv \ket{b_1}_{i,1}\cdots\ket{b_d}_{i,d}$} in the computational basis yields a bitstring that, when converted to a decimal $v$, sets the variable with index $i$ to the value with index $v+1$.
While a similar encoding of integer values has been discussed in \cite{Dominguez_2023}, here we use the encoded integers to denote value indices. This allows us to encode values beyond the set of integers, e.g., colors or gates.
In this encoding, the objective function can be written as
\begin{equation}
    \begin{split}
        \hat{H}_C & = \sum_{1 \leq i \leq n} \sum_{1\leq v \leq m} c_1(i, v) \projector{v}{i} \\ &+ \sum_{1 \leq i<j \leq n}\sum_{1 \leq v,w \leq m} c_2(i, j, v, w)  \projector{v}{i}
        \otimes \projector{w}{j}.
    \end{split}
    \label{eqn:projective_Hamiltonian}
\end{equation}
It is straightforward to check that we have $\hat{H}_C \ket{\boldsymbol{x}} = C(\boldsymbol{x}) \ket{\boldsymbol{x}}$ with this encoding.

Furthermore, if $m$ is not a power of two, we also add a penalty term in the form of
\begin{equation}
    \lambda \sum_{i=1}^{n} \sum_{v=m}^{2^d} \projector{v}{i},
\end{equation}
to penalize all states that encode indices larger than $m$. Additionally, to incorporate the penalty term \eqref{eqn:nonequal_constraint}, we can add a term of the form
\begin{equation}
    \lambda \sum_{i,j \in P} \sum_{v=1}^{m} \projector{v}{i} \otimes \projector{v}{j},
\end{equation}
to the Hamiltonian. The addition of a one-hot constraint (\equationLabel ~\eqref{eqn:one_hot_constraint}) is not required in this encoding and, as we will show, this substantially reduces the resource requirements.

For the construction of the quantum circuit, however, we need to encode the Hamiltonian in terms of Pauli matrices.
This can be done by noting that each projector-term can be written out as
\begin{equation}
    \begin{split}
        \projector{v}{i} & = \projector{v_1}{i,1} \cdots \projector{v_d}{i,d}                                              \\
                         & = \projectorExpanded{v_1}{i,1} \cdots \projectorExpanded{v_d}{i,d}                              \\
                         & = \dfrac{1}{2^d} \sum_{S \subseteq \{1, \cdots, d\}} \prod_{a \in S} (-1)^{v_a} \hat{Z}_{i, a},
    \end{split}
    \label{eqn:expanded_projector}
\end{equation}
where $v_a$ is the $a$th bit of the binary representation of $v$ and $\hat{Z}_{i,a}$ is the Pauli-Z operator acting on the qubit encoding the $a$th bit of the $i$th variable.
Note that the sum over $S$ runs over all subsets of the set $B =\{1, \cdots, d\}$, which includes the empty set. The empty set contributes a global factor of $\mathbb{I}/2^d$ to the sum.

To simplify the notation, we define the product of Pauli-Z operators for a variable $i$ and a set $S \subseteq B$ as
\begin{equation}
    \hat{Z}_{i, S} = \prod_{a \in S} \hat{Z}_{i,a}.
    \label{eqn:projector_Z}
\end{equation}
Additionally, for a set $S$ and a value $v$, we define
\begin{equation}
    R_S(v) =  \dfrac{1}{2^d} \prod_{a \in S} (-1)^{v_a}.
\end{equation}
This allows us to express the projector in a more compact form as
\begin{equation}
    \projector{v}{i} = \sum_{S \subseteq B} R_S(v) \hat{Z}_{i,S}.
    \label{eqn:projector_expanded}
\end{equation}
Substituting \eqref{eqn:projector_expanded} into the Hamiltonian \eqref{eqn:projective_Hamiltonian} yields
\begin{equation}
    \begin{split}
        \hat{H}_C & = \sum_{1 \leq i \leq n} \sum_{S \subseteq B} J_{i, S} \hat{Z}_{i, S}                                            \\
                  & + \sum_{1 \leq i < j\leq n} \sum_{S_1, S_2 \subseteq B, B} J_{i, j, S_1, S_2} \hat{Z}_{i, S_1} \hat{Z}_{j, S_2},
    \end{split}
    \label{eqn:projective_hamiltonian_expanded}
\end{equation}
where
\begin{equation}
    \begin{split}
        J_{i, S}           & = \sum_{v=1}^{m} R_S(v) c_1(i, v),                       \\
        J_{i, j ,S_1, S_2} & = \sum_{v,w=1}^{m} R_{S_1}(v) R_{S_2}(w) c_2(i,j, v, w).
    \end{split}
    \label{eqn:coefficients_hubo_hamiltonian}
\end{equation}
\equationLabel \eqref{eqn:projective_hamiltonian_expanded} is a HUBO Hamiltonian with linear, and quadratic terms and terms of order up $2 \times d$ in the Pauli-Z operators.
The coefficients $J_{i,S}$ are determined by $c_1(i, v)$ and the coefficients $J_{i, j S_1, S_2}$ by $c_2(i,j, v, w)$.
Hence, \equationLabel \eqref{eqn:coefficients_hubo_hamiltonian} provides a systematic way to find the coefficients of the HUBO Hamiltonian from the coefficients of the objective function.

In principle, the HUBO-coefficients $J_{i, S}$ and $J_{i, j, S_1, S_2}$ can be computed from \eqref{eqn:coefficients_hubo_hamiltonian}.
However, this quickly becomes \response{cumbersome from an implementation perspective, requiring careful manual tracking of all coefficients and explicit caching to avoid redundant computations}. Instead, we use a technique based on the Walsh-Hadamard transform \cite{hadfieldRepresentationBooleanReal2021}, which we present in Appendix \ref{sec:appendix_projective_to_hubo}. We also prove that calculating the HUBO-coefficients has a numerical complexity that scales polynomially with the number of variables $n$ and values $m$. For example, calculating the HUBO-coefficients for a COP with a quadratic objective function has numerical complexity \response{$\mathcal{O}(n^2 m^4)$}. Our open-source repository offers the Python package PyHUBO \cite{Koch_PyHUBO_2025}, along with detailed documentation, requirements, and installation instructions, that facilitates the computation of HUBO coefficients for COPs.

\subsection{Circuit Implementation}
\label{sec:quantum_implementation}
In the previous sections, we presented two methods that encode a COP into a Hamiltonian, ensuring that the ground state of the Hamiltonian corresponds to the optimal solution. Various techniques exist to find this ground state, such as adiabatic evolution \cite{farhiQuantumAdiabaticEvolution2001, albashAdiabaticQuantumComputation2018a}, quantum imaginary time evolution \cite{mcardleVariationalAnsatzbasedQuantum2019, beachMakingTrottersSprint2019a, loveCoolingImaginaryTime2020}, and variational methods \cite{peruzzoVariationalEigenvalueSolver2014, tillyVariationalQuantumEigensolver2022}. In this work, we focus on the QAOA \cite{farhiQuantumApproximateOptimization2014}, a variational approach well-suited for near-term quantum devices and effective across many COPs \autocite{zhouQuantumApproximateOptimization2020,bassoPerformanceLimitationsQAOA2022,blekosReviewQuantumApproximate2024,goldenNumericalEvidenceExponential2023, weidenfellerScalingQuantumApproximate2022, kurowskiApplicationQuantumApproximate2023}. QAOA is widely used in the  community and has been implemented on multiple quantum devices, making it an ideal benchmark for our encoding schemes \autocite{wang$XY$mixersAnalyticalNumerical2020,wangQuantumAlternatingOperator,sachdevaQuantumOptimizationUsing2024}.

\begin{figure}
    \centering
    \includegraphics[width=85mm]{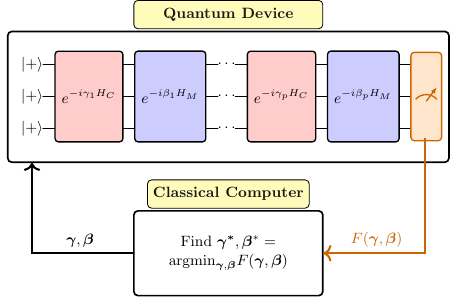}
    \caption{Schematic representation of the QAOA algorithm showing the initialization in the superposition state, the alternating application of cost and mixer unitaries across layers, followed by measurement and classical optimization of variational parameters $\boldsymbol{\gamma}$ and $\boldsymbol{\beta}$.}
    \label{fig:qaoa_circuit}
\end{figure}

QAOA operates by alternating between two types of unitary operations: the cost unitary $U_C(\gamma) = \exp(-i\gamma \hat{H}_C)$ with the cost Hamiltonian $\hat{H}_C$ as developed in the \sectionLabel (\ref{sec:one_hot_encoding}) and \sectionLabel (\ref{sec:higher_order_encoding}), and the mixer unitary $U_M(\beta) = \exp(-i\beta \hat{H}_M)$ (see figure \ref{fig:qaoa_circuit}). Below, we outline the circuit constructions for these unitaries. For a comprehensive overview of QAOA, see \cite{blekosReviewQuantumApproximate2024}.

\response{For the diagonal cost unitary, we largely follow the circuit-synthesis framework of Welch et al. for arbitrary diagonal unitaries \cite{welchEfficientQuantumCircuits2014}, and apply it here to both QUBO and HUBO Hamiltonians.}  A key observation that simplifies the circuit construction is that the cost Hamiltonian $\hat{H}_C$ for every encoding, when expressed in the computational basis, is a sum of diagonal terms. This diagonal structure \response{and, equivalently, because all Pauli-Z operators mutually commute,} implies that any two summands in the cost Hamiltonian commute with each other. This commutativity property allows us to decompose the cost unitary for both QUBO and HUBO encodings into a product of unitaries that each implement single terms $J_T \hat{Z}_T$ of the Hamiltonians \eqref{eqn:one_hot_hamiltonian} and \eqref{eqn:projective_hamiltonian_expanded}. Here, $T$ is a set of qubit indices $J_T$ is the coefficient for these qubits and, similar to \sectionLabel \ref{sec:higher_order_encoding} we use the notation $\hat{Z}_T = \prod_{t \in T} \hat{Z}_t$. Obviously, in the QUBO encoding, the set $T$ only involves at most two qubit indices, while it can involve up to $2 \times d$ qubit indices in the HUBO encoding.

To construct a circuit for the unitary $\exp(i \gamma J_T \hat{Z}_T)$ using only single- and two-qubit gates, we exploit the fact that the computational basis states $\ket{\mathbf{b}} = \bigotimes_{t \in T} \ket{b_t}_t$, with $b_t \in \{0,1\}$, form a complete basis. The action of $\exp(i \gamma J_T \hat{Z}_T)$ on such a state results in $\exp\big((-1)^{\bigoplus_{t \in T} b_t} i \gamma J_T \big) \ket{\mathbf{b}}$. We therefore design a circuit composed of single- and two-qubit gates that reproduces this transformation on computational basis states and thus implements the unitary $\exp(i \gamma J_T \hat{Z}_T)$.

\begin{figure}
    \centering
    \includegraphics[width=85mm]{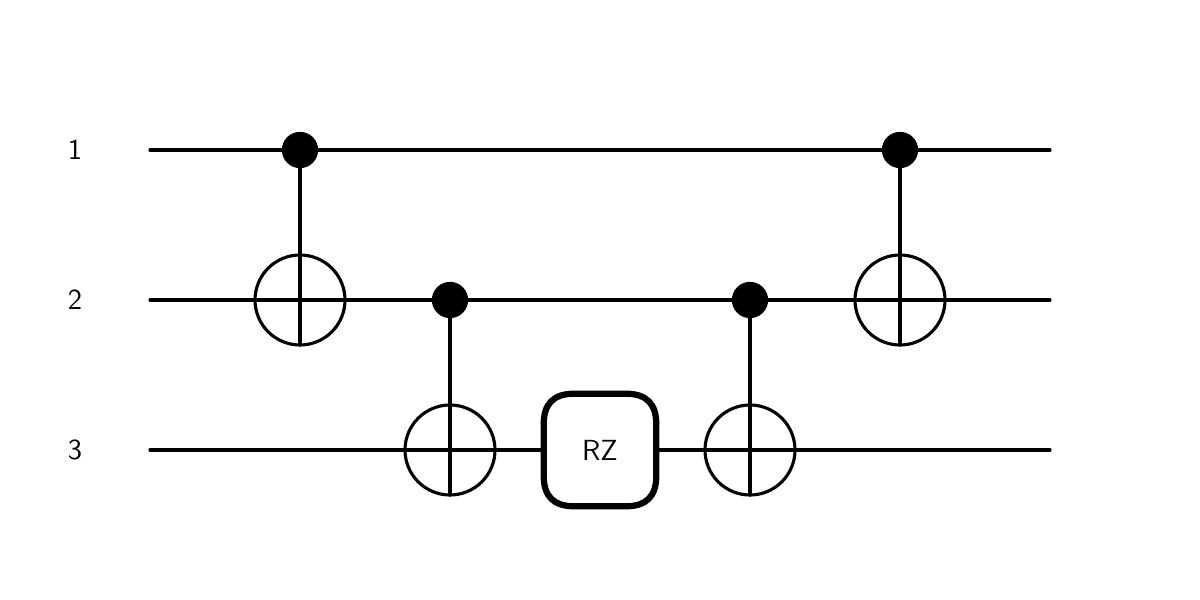}
    \caption{Circuit implementation of the cost Hamiltonian for a three-qubit term, i.e. $\exp(i \gamma J_{1,2,3} \hat{Z}_1 \hat{Z}_2 \hat{Z}_3)$. The circuit consists of a sequence of CNOT gates to entangle the qubits, such that the initial state $\ket{b_1}_1 \ket{b_2}_2 \ket{b_3}_3$ in the computational basis becomes $\ket{b_1}_1 \ket{b_1 \oplus b_2}_2 \ket{b_1 \oplus b_2 \oplus b_3}_3$ through the first chain of CNOT gates. The single qubit rotation $R_z(2\gamma J_{1,2,3})$ acts on the last qubit. After the rotation, another sequence of CNOT gates disentangles the qubits, resulting in the final state $\exp((-1)^{b_1 \oplus b_2 \oplus b_3}i \gamma J_{1,2,3})\ket{b_1}_1 \ket{b_2}_2 \ket{b_3}_3$.}
    \label{fig:three_qubit_expansion}
\end{figure}

The circuit is constructed in three main steps, as shown in Fig. \ref{fig:three_qubit_expansion}. First, a sequence of $|T| - 1$ CNOT gates is used to transform the initial state $\ket{\mathbf{b}}$ such that one qubit stores the parity $\oplus_{t \in T} b_t$ of the bitstring. Specifically, by arranging the $|T| - 1$ CNOT gates in a chain, where the $t$-th CNOT acts on the $t$-th (control) and $(t+1)$-th (target) qubits of $T$, the parity is encoded in the last qubit of the set $T$. Second, a single-qubit rotation $R_z(2\gamma J_T)$ is applied to the parity qubit, generating the phase factor $\exp(-i J_T \gamma)$ or $\exp(i J_T \gamma)$ for odd or even parity, respectively. Finally, the sequence of CNOT gates is applied in reverse order to restore the original product state, resulting in $\exp((-1)^{\oplus_{t \in T}b_t} i \gamma J_T)\ket{\mathbf{b}}$. Thus, the circuit implementing the cost unitary for a single $|T|$-th order term in the cost Hamiltonian requires $2(|T|-1)$ CNOT gates and one $R_Z$ gate.

In principle, each term in \eqref{eqn:one_hot_hamiltonian} and \eqref{eqn:projective_hamiltonian_expanded} can be implemented in this way sequentially. However, as we show in the Appendix \ref{sec:appendix_cost_hamiltonian_circuit}, constructing the cost layer $U_C(\gamma)$ for all terms of a HUBO Hamiltonian \eqref{eqn:projective_hamiltonian_expanded} in this way is not \response{the most resource-efficient approach}, and we present an circuit \response{based on the provably optimal Gray-code implementation by Welch et al. \cite{welchEfficientQuantumCircuits2014}}, which reduces the number of CNOT gates. The key idea is that parity states generated for one term of the Hamiltonian can often be leveraged to efficiently construct the parity states needed for other terms, thereby minimizing redundant operations. Consequently, as we show in Appendix \ref{sec:appendix_derivation_scaling}, the asymptotic CNOT and $R_Z$ gate counts \response{for the Gray-code implementation scale as $O(n^2 m^2)$, whereas the sequential implementation scales as $O(n^2 m^2 \log(m))$}.

For the mixer Hamiltonian $\hat{H}_M$, we employ the standard Pauli-X mixer
\begin{equation}
    \hat{H}_M = \sum_{i} \hat{X}_i,
\end{equation}
which acts on all qubits of the circuit. This choice ensures that the mixer unitary can efficiently explore the computational basis states by creating a superposition that allows transitions between different solution candidates. Furthermore, the mixer unitary can be implemented using a simple layer of single-qubit $R_x$ rotation gates.

To optimize the variational parameters $\boldsymbol{\gamma}$ and $\boldsymbol{\beta}$, we apply gradient descent methods with the initialization scheme of Zhou et al. \cite{zhouQuantumApproximateOptimization2020}, which improves convergence and solution quality over random initialization. Gradients of the expectation value with respect to variational parameters are efficiently computed via implicit differentiation \cite{jaxopt_implicit_diff}, avoiding the computational overhead and inaccuracy of finite-difference methods. Quantum circuits and optimization routines are implemented using PennyLane \cite{bergholmPennyLaneAutomaticDifferentiation2022}, enabling automatic differentiation and GPU acceleration for efficient optimization.

The Lagrange multipliers $\lambda$ for the different penalty terms in the objective function are chosen iteratively before running the QAOA benchmarking.
Only after identifying $\lambda$ values that ensure the QAOA algorithm consistently finds feasible solutions respecting all constraints do we proceed to run and evaluate the algorithm’s performance using these selected penalty parameters. This iterative approach ensures that the penalty terms are effective in guiding the QAOA algorithm towards feasible solutions while still allowing for exploration of the solution space.

The quality of QAOA solutions is typically evaluated via the expectation value of the cost Hamiltonian. For COPs, however, feasibility must also be considered. Hence, we employ a metric similar to the one introduced in \cite{schulzGuidedQuantumWalk2024}, here referred to as the approximation ratio, defined as
\begin{equation}
    A = 1 - \sum_{\mathbf{b} \in B_{\text{feas}}}  r(\mathbf{b}) \abs{\bra{\psi(\boldsymbol{\gamma}, \boldsymbol{\beta})}\ket{\mathbf{b}}}^2,
    \label{eqn:approximation_ratio}
\end{equation}
where $\{\ket{\mathbf{b}}| \mathbf{b} \in B_{\text{feas}} \}$ is the set of computational basis states satisfying all constraints and
\begin{equation}
    r(\mathbf{b}) = \frac{C_{\text{max}} - C(\mathbf{b})}{C_{\text{max}} - C_{\text{min}}},
\end{equation}
rescales
$C(\mathbf{b})$ to the interval $[0,1]$. Here, $C_{\text{min}}$ and $C_{\text{max}}$ are the minimal and maximal values of the objective function, determined using commercial solvers for the benchmark instances. The approximation ratio is particularly suitable for our benchmarks, as it immediately accounts for the treatment of invalid solutions as equivalent to maximum cost states. This metric combines solution utility, feasibility, and objective value, and can be applied to all three COPs discussed in the results section.

If the quantum state $\ket{\psi (\boldsymbol{\gamma}, \boldsymbol{\beta})}$ corresponds to the optimal solution, then $A=0$. If it yields only infeasible or maximal-cost solutions, $A=1$. Thus, $A$ can be interpreted as the expectation value of a normalized cost Hamiltonian, where feasible eigenstates have eigenvalues $[0,1]$, while infeasible eigenstates can only have eigenvalue 1. Lower values of $A$ indicate higher-quality solutions found by QAOA.

Since the set of feasible bitstrings \response{$B_{\text{feas}}$} grows exponentially with problem size, evaluation of the approximation ratio becomes numerically taxing even for modestly sized COPs. To address this, we approximate \equationLabel \eqref{eqn:approximation_ratio} using a sampling-based method. Further details on this approach and other numerical techniques used for benchmarking are provided in Appendix~\ref{sec:appendix_details_qaoa}.

\section{Results}
\label{sec:results}

\begin{table}
    \centering
    \begin{tabular}{lccc}
        \toprule
             & \# Qubits                 & \# CNOTs     & \# $R_Z$     \\
        \midrule
        QUBO & $n m$                     & $O(n^2 m^2)$ & $O(n^2 m^2)$ \\
        HUBO & $n \lceil log_2(m)\rceil$ & $O(n^2 m^2)$ & $O(n^2 m^2)$ \\
        \bottomrule
    \end{tabular}
    \caption{Exact qubit requirements and analytical scaling for CNOT and $R_Z$ gates per QAOA layer in both encodings, where $n$ is the number of variables and $m$ the number of values per variable. \response{These scalings are based on worst case assumptions (see Appendix \ref{sec:appendix_derivation_scaling})}. In practice, gate counts are often lower (see Table~\ref{tab:resources_per_layer}).}
    \label{tab:scaling_table}
\end{table}
\begin{table}
    \centering
    \begin{tabular}{lccc}
        \toprule
                  & \# Qubits & \# CNOTS & \# $R_Z$ \\
        \midrule
        GAP-QUBO  & $20$      & $140$    & $90$     \\
        GAP-HUBO  & $10$      & $68$     & $27$     \\
        MkCS-QUBO & $20$      & $132$    & $86$     \\
        MkCS-HUBO & $10$      & $90$     & $27$     \\
        IP-QUBO   & $16$      & $120$    & $76$     \\
        IP-HUBO   & $8$       & $38$     & $31$     \\
        \bottomrule
    \end{tabular}
    \caption{Quantum Resources required per QAOA layer to encode the GAP instance introduced in \figureLabel \ref{fig:objective_values_gap}, an MkCS instance illustrated in \figureLabel \ref{fig:objective_values_graph_coloring} and an IP instance shown in \figureLabel \ref{fig:objective_values_ip}.}
    \label{tab:resources_per_layer}
\end{table}

To evaluate resource efficiency and solution quality, we analyze three complementary metrics across all benchmarks: (i) the number of qubits required to represent the problem, (ii) the gate count per QAOA layer after circuit compilation (i.e.
\# CNOTS, \# $R_Z$), and (iii) the total number of gates needed to achieve target approximation ratios ($A$).

For qubit requirements, a COP with $n$ variables each taking $m$ values requires $n \times m$ qubits under QUBO, but only $n \times \lceil \log_2(m) \rceil$ under HUBO.
Both encodings grow linearly in $n$, however, the growth in $m$ is linear for QUBO but logarithmic for HUBO, yielding exponential savings when variables admit many values. This advantage is particularly critical in problems such as the GAP, where each flight-variable can assume a large number of airport-gate-values~\cite{bourasAirportGateAssignment2014}.

For gate counts per QAOA layer, we compile the cost unitary circuits to contain only rotation and CNOT gates, and report those as our primary resources. Both encodings scale quadratically in gate count (see Table~\ref{tab:scaling_table}). In practice, HUBO consistently compiles to fewer gates (Table~\ref{tab:resources_per_layer}), often by large margins, for instance, a 69\% \response{per-layer} reduction in CNOT gates for the considered IP instance. \response{These savings can largely be attributed to the penalty overhead from the QUBO encoding (for details see Appendix \ref{sec:oh_resource_requirements})}. It should be noted that Hadamard gates for state initialization and $R_X$ gates for the mixer are not reported separately since their counts equal the number of qubits.

Finally, while increasing the number of QAOA layers is known to improve solution quality~\cite{farhiQuantumApproximateOptimization2014}, feasibility on real hardware depends on total gate counts rather than per-layer counts alone. In what follows, we will show that both the per-layer gate requirements and the convergence rate toward the optimum vary significantly with the chosen encoding. Since no analytical bounds exist for the minimum depth needed to reach a given solution threshold, we rely on empirical benchmarking. Here, the most notable advantage of HUBO over QUBO is evident, with HUBO achieving savings of up to 97.1\% in CNOT counts for IP benchmarks. In the following subsections, we therefore compare QUBO and HUBO across three representative COPs, reporting solution quality at fixed depths, convergence behavior as a function of gate count, and total resources needed to achieve specified solution thresholds. All averages are over 100 independently optimized QAOA parameters, with error bars representing the standard deviation, as detailed in Appendix \ref{sec:appendix_details_qaoa}.

\subsection{Gate Assignment Problem}
\begin{figure}
    \centering
    \includegraphics[width=85mm]{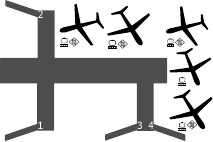}

    \includegraphics[width=85mm]{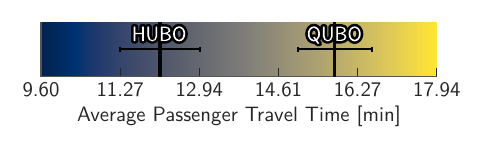}

    \caption{Upper panel: Airport‑Gate layout for the benchmark instance with 4 airport‑gates and 5 flights. Lower panel: The range of feasible objective values for total passenger time is shown, including the minimum (blue) and maximum (yellow) obtained by a commercial solver. The average objective values achieved by the QAOA algorithm with QUBO and HUBO encodings (10 layers) are shown for comparison.}
    \label{fig:objective_values_gap}
\end{figure}
Using a GAP instance with five flights ($n=5$) and four airport-gates ($m=4$), we benchmark both QUBO-QAOA and HUBO-QAOA, with ten layers. Each layer contains gate counts as reported in Table \ref{tab:resources_per_layer}.  The GAP instance considered has $4$ flights with transfer passengers, resulting in $24$ nonzero quadratic terms (see Equation~\eqref{eqn:gap_quadratic_term}) in the classical objective function. Additionally, $3$ flights have gate conflicts with $2$ other flights each, while $2$ flights have gate conflicts with $1$ other flight each. The motivation for this scenario, as well as further details such as passenger numbers and walking times, are provided in Appendix~\ref{sec:appendix_gap_benchmark_details}. An illustration of the problem is shown in Figure~\ref{fig:objective_values_gap}.

The best feasible solution, determined by a commercial solver, yields an average passenger travel time of $9.6$ minutes, while the worst feasible assignment totals $17.9$ minutes. Within this range of solutions, HUBO-QAOA achieves an average objective value of $12.1$ minutes, while QUBO-QAOA's average objective value is $15.8$ minutes, demonstrating a substantial solution quality improvement with HUBO for the same number of QAOA layers.

\begin{figure}
    \centering
    \includegraphics[width=85mm]{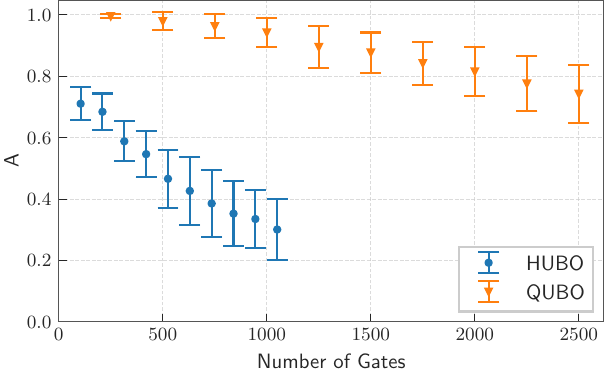}
    \caption{Average approximation ratio (Eq.~\eqref{eqn:approximation_ratio}) versus total gate count for HUBO and QUBO encodings on the 4-airport-gate, 5-flight problem instance. The gate count includes all circuit gates used: CNOT, $R_Z$, $R_X$, and Hadamard, the latter applied only once for state initialization (10 for HUBO, 20 for QUBO). For each encoding, the number of gates per QAOA layer is fixed, so increasing the number of layers results in a uniform increment in total gates, producing evenly spaced datapoints along the x-axis. Each point corresponds to a fixed number of QAOA layers, ranging from 1 to 10.}
    \label{fig:approximation_ratio_gap}
\end{figure}
Furthermore, HUBO-QAOA achieves lower approximation ratios using fewer quantum resources. In \figureLabel \ref{fig:approximation_ratio_gap}, we compare the average approximation ratio as a function of total gate count, varying the number of QAOA layers from $1$ to $10$. Because the same number of gates is added per QAOA layer, the datapoints are evenly spaced along the x-axis. For any fixed number of QAOA layers, the HUBO encoding requires fewer gates in total. More importantly, for any given total gate count, HUBO-QAOA consistently attains much better approximation ratios. For example, with approximately 1040 total gates (1060 for HUBO, 1020 for QUBO), the approximation ratio achieved with QUBO is more than 3 times worse than that achieved with HUBO.

\begin{figure}
    \centering
    \includegraphics[width=85mm]{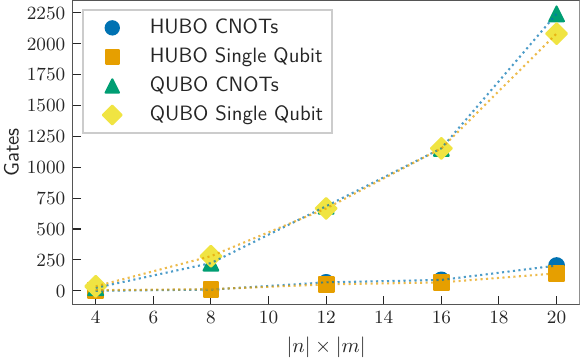}
    \caption{Total number of CNOT and single-qubit gates required to attain an approximation ratio of $0.50$, plotted against the problem size of the GAP (number of airport-gates fixed at $4$). Problem size is defined as the product of the number of flights and the number of gates and directly corresponds to the qubit count in the QUBO encoding. CNOT and single-qubit gates are shown separately, reflecting their differing experimental costs. Across all problem sizes, the HUBO encoding achieves the target approximation ratio using substantially fewer CNOT and single-qubit gates than the QUBO encoding.}
    \label{fig:qaoa_scaling_gap}
\end{figure}
HUBO-QAOA consistently demonstrates resource savings over QUBO-QAOA as the problem size varies. \figureLabel\ref{fig:qaoa_scaling_gap} summarizes the number of gates required to achieve an approximation ratio of at least $0.50$ for both encodings, as the number of flights increases from $1$ to $5$ with the number of airport-gates fixed at $4$. The smaller GAP instances are generated by repeatedly deleting flight assignments from the instance described in \figureLabel\ref{fig:objective_values_gap}. In this scaling approach, the problem size corresponds directly to the number of qubits required by the QUBO encoding. The figure separately reports the number of CNOT and single-qubit gates, since CNOT gates are typically more experimentally demanding. Across all evaluated problem sizes, the HUBO encoding reduces the number of required CNOT gates between $90.0\%$ and $100\%$ (the single flight GAP instance in the HUBO encoding doesn't require CNOT gates) and the number of $R_Z$ gates between $86.1\%$ and $95.3\%$ compared to the QUBO encoding. These results highlight that substantial quantum resource efficiency gains with the HUBO approach persist across varying problem sizes.

\subsection{Maximum k-Colorable Subgraph Problem}
\begin{figure}
    \centering
    \includegraphics[width=85mm]{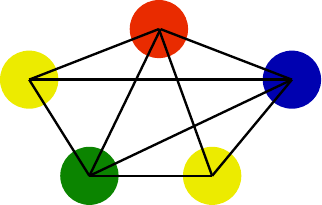}
    \includegraphics[width=85mm]{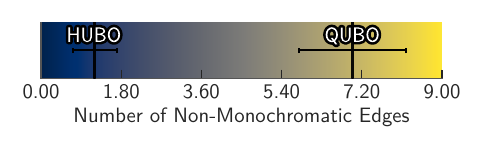}
    \caption{Upper panel: Graph with five vertices and nine edges used as a benchmark instance for the MkCS problem, colored with four colors. The displayed coloring corresponds to the optimal solution with zero monochromatic edges. Lower panel: Range of feasible objective values and average number of monochromatic edges identified by HUBO-QAOA and QUBO-QAOA for a fixed number of layers ($3$).}
    \label{fig:objective_values_graph_coloring}
\end{figure}
We benchmark an MkCS problem with $|V|=n=5$ vertices and $k=m=4$ colors on the graph illustrated in \figureLabel \ref{fig:objective_values_graph_coloring}, using a QAOA circuit with three layers. The objective is to minimize the number of monochromatic edges. For this instance, an optimal coloring results in zero monochromatic edges, while assigning the same color to all vertices yields a maximum objective value of nine. The range of feasible values and the average performance of each encoding are shown in the lower panel of \figureLabel \ref{fig:objective_values_graph_coloring}. HUBO-QAOA finds, on average, $1.2$ monochromatic edges, compared to an average of $6.9$ for QUBO-QAOA.

\begin{figure}
    \centering
    \includegraphics[width=85mm]{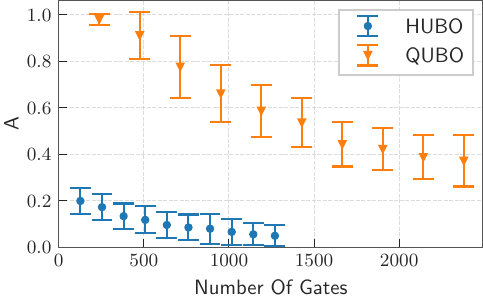}
    \caption{Average approximation ratio achieved by QAOA as a function of total gate count for HUBO and QUBO encodings. Results are shown for the same graph with five vertices and nine edges as in \figureLabel\ref{fig:objective_values_graph_coloring}.
        The number of gates required for each layer can be found in Table \ref{tab:resources_per_layer}. Data points are evenly spaced as gate counts scale proportionally with the number of QAOA layers.}
    \label{fig:approximation_ratio_graph_coloring}
\end{figure}
HUBO-QAOA consistently achieves better approximation ratios than QUBO-QAOA while using fewer quantum resources. \figureLabel\ref{fig:approximation_ratio_graph_coloring} shows the average approximation ratio obtained by QAOA as a function of total gate count for both encodings.
Using the gate numbers reported in Table~\ref{tab:resources_per_layer}, HUBO-QAOA always reaches approximation ratios below $0.18$. In contrast, even with the largest considered number of QAOA layers (10), the QUBO-QAOA algorithm only attains an average approximation ratio of $0.37$. Each QUBO-QAOA layer requires 238 gates, compared to only 127 gates for HUBO-QAOA. This substantial difference in resource demands can be attributed to the penalty overhead required by the QUBO encoding.

\begin{figure}
    \centering
    \includegraphics[width=85mm]{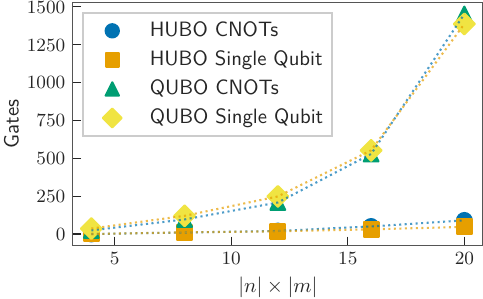}
    \caption{Scaling of quantum resource requirements for MkCS instances. The number of CNOT and $R_Z$ gates required to achieve an approximation ratio of $0.20$ is shown as a function of problem size (number of vertices times number of colors, with $m=4$ fixed). The HUBO encoding consistently requires substantially fewer gates than QUBO across all tested sizes, with the difference increasing for larger instances.}
    \label{fig:qaoa_scaling_graph_coloring}
\end{figure}
We find that the quantum resource advantage of HUBO over QUBO persists across all tested MkCS problem sizes. As summarized in \figureLabel\ref{fig:qaoa_scaling_graph_coloring}, the number of gates required to achieve an approximation ratio of $0.20$ increases much more rapidly for QUBO than for HUBO as problem size grows. Problem size is defined as the product of the number of vertices and the number of colors in the graph, which matches the qubit requirement for the QUBO encoding. For these benchmarks, the number of colors is fixed at $k=4$ and the number of vertices varies from $1$ to $5$. The smaller graphs are generated by deleting vertices from the instance in \figureLabel\ref{fig:objective_values_graph_coloring}. CNOT and single-qubit gate requirements are reported separately (note that the one-vertex instance is solved trivially by HUBO and doesn't require any CNOT or RZ gates). Across all evaluated problem sizes, the HUBO encoding reduces the number of required CNOT gates by 89.6 -- 100\% and the number of $R_Z$ gates by 90.8 -- 100\% compared to the QUBO encoding.

\subsection{Integer Programming}
\begin{figure}
    \centering
    \includegraphics[width=85mm]{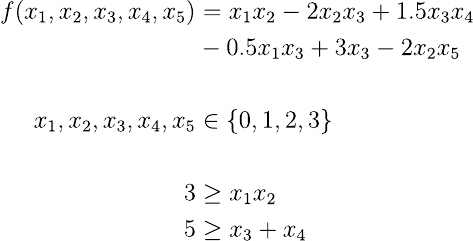}
    \includegraphics[width=85mm]{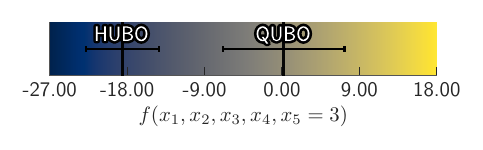}
    \caption{Upper panel: Structure of the IP instance considered, with $n=5$ variables ($m=4$ values each), two linear constraints, and both linear and quadratic objective function terms. This instance is referenced here for later HUBO-QAOA benchmarking, as QUBO-QAOA is not computationally feasible for this size.  Lower panel: Range of feasible objective values for the reduced IP instance with $x_5=3$ fixed. The optimal ($-27$) and maximal ($18$) values are shown. Average objective values attained by HUBO-QAOA and QUBO-QAOA after nine layers are indicated for comparison.}
    \label{fig:objective_values_ip}
\end{figure}
To provide context for the subsequent analyses, the upper panel of \figureLabel\ref{fig:objective_values_ip} illustrates the IP instance considered in this work. This instance consists of $n=5$ variables, each with $m=4$ possible values, two linear constraints (each depending on two variables), and an objective function containing both linear and quadratic terms. The minimum feasible objective value is $-27$, which is achieved by the assignment $x_1=0$, $x_2=3$, $x_3=3$, $x_4=0$, and $x_5=3$. This instance is used to benchmark HUBO-QAOA; however, QUBO-QAOA was not computationally feasible at this problem size due to excessive computational resource requirements.

To enable a direct comparison between the two encodings, we benchmark QAOA on a reduced version of the problem with $x_5=3$ fixed. The lower panel of \figureLabel\ref{fig:objective_values_ip} displays the range of feasible objective values for this simplified instance, which retains the optimum ($-27$) and maximum ($18$) values of the original. For $9$ QAOA layers, HUBO-QAOA yields an average objective value of $-18.55$, while QUBO-QAOA produces an average of $0.21$.

\begin{figure}
    \centering
    \includegraphics[width=85mm]{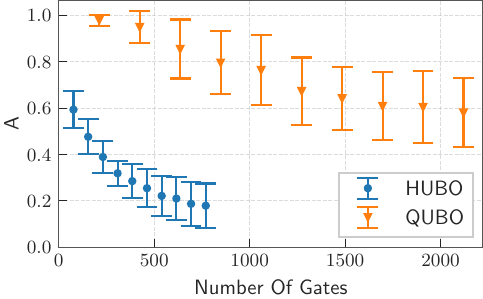}
    \caption{Average approximation ratio versus total gate count for QAOA applied to the reduced IP instance, comparing HUBO and QUBO encodings. HUBO-QAOA consistently achieves higher approximation ratios at lower quantum resource cost than QUBO-QAOA. For example, to reach a target approximation ratio of $0.60$, HUBO-QAOA and QUBO-QAOA require $85$ and $2136$ gates, respectively.}
    \label{fig:approximation_ratio_ip}
\end{figure}
Additionally, HUBO-QAOA demonstrates significantly greater resource efficiency and superior approximation ratio compared to QUBO-QAOA across all layer settings, as shown in \figureLabel\ref{fig:approximation_ratio_ip}. The per-layer CNOT and $R_Z$ gate counts for each encoding are provided in Table~\ref{tab:resources_per_layer}. Evidently, QUBO-QAOA requires over three times more CNOT gates and more than twice as many $R_Z$ gates per layer compared to HUBO-QAOA.

For a targeted approximation threshold, HUBO-QAOA requires substantially fewer gates compared to QUBO-QAOA. For instance, to reach an approximation ratio of $0.60$, QUBO-QAOA requires $10$ layers, while HUBO-QAOA achieves the same result with only $1$ layer. When accounting for both the higher resources per layer and the greater number of layers required, QUBO-QAOA, compared to HUBO-QAOA, consumes 31.6 times more CNOT gates (1200 vs. 38) and 24.5 times more $R_Z$ gates (760 vs. 31).

\begin{figure}
    \centering
    \includegraphics[width=85mm]{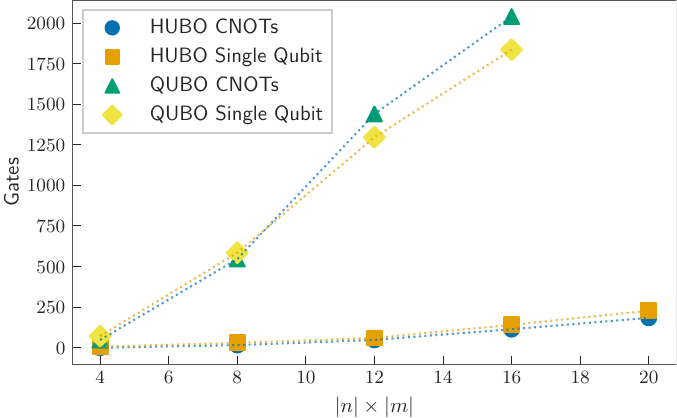}
    \caption{Number of gates required to achieve an approximation ratio of $0.30$ as a function of IP problem size, comparing HUBO and QUBO encodings. Both CNOT and single-qubit gate requirements are shown. Problem size is modified by changing the number of variables in the IP problem. HUBO-QAOA consistently achieves the target approximation ratio with dramatically fewer gates than QUBO-QAOA, and remains tractable even for the largest instances where QUBO-QAOA cannot be executed due to computational resource constraints.}
    \label{fig:qaoa_scaling_ip}
\end{figure}
The resource advantage of HUBO-QAOA over QUBO-QAOA remains robust across all benchmarked IP problem sizes. \figureLabel\ref{fig:qaoa_scaling_ip} reports the number of gates required to reach an approximation ratio of $0.30$ for different problem sizes, with CNOT and single-qubit gate requirements shown separately. Problem size is varied by starting from the largest instance (as shown in \figureLabel\ref{fig:objective_values_ip}, with $n=5$ variables, and $m = 4$ values) and generating progressively smaller instances by fixing variables to their optimal assignments. This approach preserves the minimal objective value while producing a sequence of simpler problems, resulting in increments of $4$ qubits per data point.

Across all benchmarked problem sizes, the HUBO encoding reduces CNOT requirements by 94.4 -- 100\% and the number of $R_Z$ gates by 91.6 -- 95.3\% compared to QUBO (again, the one variable instance doesn't require any CNOT gates in the HUBO formulation). These results, together with those from earlier figures, demonstrate the persistent resource efficiency advantage of HUBO-QAOA over QUBO-QAOA throughout the tested range of IP problem sizes.

\section{Conclusion}
We showed that using HUBO formulations
\response{can substantially reduce} resource bottlenecks that often constrain the practical quantum optimization of COPs. Although one might assume that higher-order interactions would increase quantum hardware demands, our results on three representative problems reveal the opposite. HUBO encoding requires \response{significantly} fewer qubits compared to QUBO, and more importantly, leads to substantial reductions in the number of CNOT and single-qubit gates. For each problem class, we observe a reduction of CNOT gates by 89.6 -- 100\%, which makes experimental realization more feasible for current and near-term quantum devices.

In addition to these resource savings, we have developed a general framework for constructing HUBO models, together with the open-source Python package \cite{Koch_PyHUBO_2025}.
This equips the community with a framework to efficiently formulate COPs on quantum hardware, helping to mitigate current resource limitations, and, for quadratic problems, ensures that the Hamiltonian construction scales polynomially with problem size.

\response{Our results suggest that HUBO encodings can be a valuable tool for reducing resource demands in quantum optimization. As quantum technology advances and problem sizes scale up, such resource-efficient methods will be increasingly important for} enabling practical applications in industry and research.

\backmatter

\section*{Acknowledgements}

For this work, the HPC-cluster Hummel-2 at University of Hamburg was used. The cluster was funded by Deutsche Forschungsgemeinschaft (DFG, German Research Foundation) – 498394658.
Für diese Arbeit wurde das HPC-Cluster Hummel-2 an der Universität Hamburg genutzt, das durch die Deutsche Forschungsgemeinschaft (DFG) – 498394658 – gefördert wurde. The authors also thank Felix Herbort for his technical support and Robert Glöckner and Shaham Jafarpisheh for helpful discussions.
\response{The authors also thank Igor Gaidai for a thorough reading of the manuscript and valuable feedback.}

\section*{Declarations}

\subsection*{Funding}
Open Access funding enabled and organized by Projekt DEAL. The authors gratefully acknowledge the funding provided by the Hamburgische Investitions- und Förderbank for the project "Efficient Quantum Algorithms for Aviation". SP and DJ acknowledge support from the Hamburg Quantum Computing Initiative (HQIC) project EFRE. The EFRE project is co-financed by ERDF of the European Union and by “Fonds of the Hamburg Ministry of Science, Research, Equalities and Districts (BWFGB)”. DJ acknowledges support from the Cluster of Excellence `Advanced Imaging of Matter' of the Deutsche Forschungsgemeinschaft (DFG) - EXC 2056 - project ID 390715994, the European Union’s Horizon Programme (HORIZON-CL42021-DIGITALEMERGING-02-10) Grant Agreement 101080085 QCFD, and DFG project ‘Quantencomputing mit neutralen Atomen’(JA 1793/1-1, Japan-JST-DFG-ASPIRE 2024). DJ acknowledges funding from the Federal Ministry of Research, Technology, and Space (BMFTR) under the grant BeRyQC.
\subsection*{Conflict of interest/Competing interests}
The authors declare that they have no competing interests.
\subsection*{Data availability}
The datasets generated and/or analysed during the current study are available in the UHH Forschungsdatenrepositorium repository, https://doi.org/10.25592/uhhfdm.18252.
\subsection*{Author contribution}
FK developed the required codes for the calculations and plotted the graphs. FK and SP analyzed the data and wrote the manuscript. RM, JD, and DJ provided comments for improving the manuscript. JD and DJ are the supervisors of the project.

\begin{appendices}

    \section{Mathematical Foundation of HUBO Encoding}
\label{sec:appendix_projective_to_hubo}
While in principle it is possible to find the HUBO coefficients using \equationLabel \eqref{eqn:coefficients_hubo_hamiltonian}, it quickly becomes cumbersome and numerically expensive to iterate over all required subsets.
Instead, we adopt a more efficient, less error-prone, and clearer method to determine the HUBO coefficients. The following derivation is inspired by \cite{hadfieldRepresentationBooleanReal2021}, where the authors show that finding the coefficients of a HUBO Hamiltonian from a Boolean function amounts to Fourier transforming it.

We first note that a single qubit projector can be written out as
\begin{equation}
    \begin{pmatrix}
        c_0 \\ c_1
    \end{pmatrix}
    \begin{pmatrix}
        \projector{0}{} \\
        \projector{1}{}
    \end{pmatrix}
    = \dfrac{1}{2}
    \begin{pmatrix}
        c_0+c_1 \\ c_0-c_1
    \end{pmatrix}
    \begin{pmatrix}
        \mathbb{I} \\ \hat{Z}
    \end{pmatrix}.
    \label{eqn:single_projector_expanded}
\end{equation}
In the following, we will call
$\begin{pmatrix}
        \projector{0}{} \\
        \projector{1}{}
    \end{pmatrix}$,
$p$-basis and
$\begin{pmatrix}
        \mathbb{I} \\ \hat{Z}
    \end{pmatrix}$,
$j$-basis and rewrite \equationLabel \eqref{eqn:single_projector_expanded} as
\begin{equation}
    \begin{pmatrix}
        c_0 \\ c_1
    \end{pmatrix}_p
    =
    \dfrac{1}{2}
    \begin{pmatrix}
        c_o+c_1 \\ c_0-c_1
    \end{pmatrix}_j
    =
    \hat{H} \begin{pmatrix}
        c_0 \\ c_1
    \end{pmatrix}_j
    \coloneq
    \begin{pmatrix}
        J_0 \\ J_1
    \end{pmatrix}_j,
\end{equation}
where $\hat{H}$ is the Hadamard matrix
providing a transformation from the $p$-basis to the $j$-basis.

Generalizing this idea to an $d$-qubit projector, the coefficient $c_{i_1 i_2 \cdots i_d}$ of the projector $\projector{i_1 i_2 \cdots i_d}{}$ in the $p$-basis is transformed to the $j$-basis using the Hadamard matrix $\hat{H}^{\otimes d}$, which is the tensor product of $d$ Hadamard matrices. This can be seen by noting that the projector $\projector{i_1 i_2 \cdots i_d}{}$ can be written as the tensor product of the single-qubit projectors $\projector{i_1}{}\otimes \projector{i_2}{}\otimes \cdots \otimes \projector{i_d}{}$, each being transformed into the $j$-basis by a Hadamard matrix, hence the transformation for the entire $d$-qubit projector is done by the tensor product of Hadamard matrices for all single qubit coefficients.

We use the notation $c_{i_1 i_2 \cdots i_d}$ for the coefficient of the projector $\projector{i_1 i_2 \cdots i_d}{}$ in the $p$-basis and the coefficient $J_{i_1 i_2 \cdots i_d}$ in the $j$-basis encodes the interaction strength of the HUBO term \response{ $\prod_{j=1}^d \hat{Z}_j^{i_j}$}.
Using this notation, we can write the transformation from the $p$-basis to the $j$-basis as
\begin{equation}
    \begin{pmatrix}
        c_{0\cdots 0 0} \\
        c_{0\cdots 0 1} \\
        \vdots          \\
        c_{1\cdots 1 0} \\
        c_{1\cdots 1 1}
    \end{pmatrix}_p
    =
    \hat{H}^{\otimes d}
    \begin{pmatrix}
        c_{0\cdots 0 0} \\
        c_{0\cdots 0 1} \\
        \vdots          \\
        c_{1\cdots 1 0} \\
        c_{1\cdots 1 1}
    \end{pmatrix}_j
    = \begin{pmatrix}
        J_{0\cdots 0 0} \\
        J_{0\cdots 0 1} \\
        \vdots          \\
        J_{1\cdots 1 0} \\
        J_{1\cdots 1 1}
    \end{pmatrix}_j.
\end{equation}
For example, a quadratic COP has the HUBO model
\begin{equation}
    \hat{H} = \sum_{1 \leq i \leq n} \sum_{1\leq k \leq m} c_1(i, k) \projector{k}{i} + \sum_{1 \leq i<j \leq n}\sum_{1 \leq k,l \leq m} c_2(i, j, k, l)  \projector{k}{i} \otimes \projector{l}{j}.
\end{equation}
We write the linear coefficients for each variable $i$ in a vector $\boldsymbol{c}(i) = (c_1(i, 1), c_1(i, 2), \ldots, c_1(i, m))^T$ and apply
$\hat{H}^{\otimes d}$ to each vector $\boldsymbol{c}_1(i)$ to obtain the coefficients in the $j$-basis $\boldsymbol{J}_i = \hat{H}^{\otimes d} \boldsymbol{c}_1(i)$. Note that because $2^d = m$, the Hadamard transformation is an $m\times m$ matrix, hence finding the linear coefficients of all $n$ variables in the $j$-basis has a computational complexity of $n$ matrix multiplications on an $m$-dimensional vector, hence \response{$O(n m^2)$}. Similarly, we can find the quadratic coefficients by applying the Hadamard transform to the vector $\boldsymbol{c}_{ij} = (c_2(i, j, 1, 1), c_2(i, j, 1, 2), \ldots, c_2(i, j, m, m))^T$ for each pair of variables $(i,j)$ and obtain the linear coefficients in the $j$-basis $\boldsymbol{J}_{ij} = \hat{H}^{\otimes d} \boldsymbol{c}_{ij}$. Calculating the quadratic coefficients for all pairs of variables amounts to a computational complexity of $n^2$
(since there are $\binom{n}{2}$ combinations of variables) and with the vector $\boldsymbol{c}_{i,j}$ being of size $m^2$ the computational complexity for finding the quadratic coefficients is \response{$O(n^2 m^4)$}. In general, for a COP of $k$th order with $n$ variables and $m$ values per variable, the computational complexity for finding the coefficients in the $j$-basis is \response{$O(\binom{n}{k} m^{2k})$}.

This formalism forms the basis of the PyHUBO Python package \cite{Koch_PyHUBO_2025}, which is made available as part of this publication.

    \section{Cost-Efficient Circuit Implementation of HUBO Hamiltonians}
\label{sec:appendix_cost_hamiltonian_circuit}
\begin{figure}[t!]
  \centering
  \includegraphics[width=85mm]{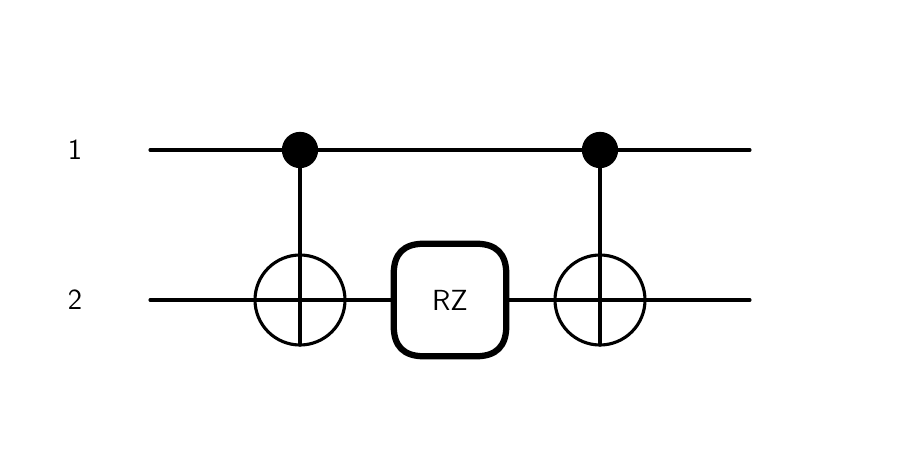}
  \includegraphics[width=85mm]{three_qubit_cost_hamiltonian.pdf}
  \caption{Circuit implementation of the cost unitary for two-qubit and three-qubit terms, i.e. $\exp(i \gamma J_{1,2} \hat{Z}_1 \hat{Z}_2)$ and $\exp(i \gamma J_{1,2,3} \hat{Z}_1 \hat{Z}_2 \hat{Z}_3)$, respectively. The circuit consists of a sequence of CNOT gates to entangle the qubits, such that the initial states $\ket{b_1}_1 \ket{b_2}_2$ and $ \ket{b_1}_1 \ket{b_2}_2 \ket{b_3}_3$ in the computational basis become $\ket{b_1}_1 \ket{b_1 \oplus b_2}_2$ and $\ket{b_1}_1 \ket{b_1 \oplus b_2}_2 \ket{b_1 \oplus b_2 \oplus b_3}_3$ through the first chain of CNOT gates. The single qubit rotation $R_Z(2\gamma J_{0,1})$ and $R_Z(2\gamma J_{0,1,2})$ acts on the last qubit. After the rotation, another sequence of CNOT gates disentangles the qubits, resulting in the final state $\exp((-1)^{b_1 \oplus b_2}i \gamma J_{1,2})\ket{b_1}_1 \ket{b_2}_2$ and $\exp((-1)^{b_1 \oplus b_2 \oplus b_3}i \gamma J_{1,2,3})\ket{b_1}_1 \ket{b_2}_2 \ket{b_3}_3$.}
  \label{fig:exponentiated_cost}
\end{figure}
An optimized circuit implementation of a HUBO Hamiltonian can be achieved using a Gray code approach, \response{introduced in \autocite{verchereOptimizingVariationalCircuits2023a} and \autocite{welchEfficientQuantumCircuits2014}, which we largely follow.} This circuit implementation has been successfully employed to problems like the traveling salesperson problem \autocite{glosSpaceefficientBinaryOptimization2022} and the circuit representation of boolean functions \autocite{amyCNOTcomplexityCNOTPHASECircuits2018}.

To understand this approach, we examine the effect of the circuit introduced in \sectionLabel (\ref{sec:quantum_implementation}) on a computational basis state $\ket{\mathbf{b}} = \otimes_{t\in T} \ket{b_t}_t$ for a set of qubit indices $T$. To construct a circuit for the term $\exp(i J_T \hat{Z}_T)$, we create entangled states where one qubit encodes the parity $\oplus_{t \in T} b_t$. \response{The $R_Z$ gate implements $\exp(-i t\hat{Z}/2 )$, so} acting with an $R_Z$ gate on this qubit results in a phase factor $\exp(i \gamma J_T)$ if $\oplus_{t \in T} b_t = 1$ and $\exp(-i \gamma J_T)$ if $\oplus_{t \in T} b_t = 0$. This is exactly how $\exp(i J_T \hat{Z}_T)$ acts on a computational basis state. We illustrate the circuit in \figureLabel \ref{fig:exponentiated_cost}, where we show the circuit diagrams to implement the terms $\exp(i \gamma J_{1,2} \hat{Z}_1 \hat{Z}_2)$ and $\exp(i \gamma J_{1,2,3} \hat{Z}_1 \hat{Z}_2 \hat{Z}_3)$. \response{Each term $\exp(-i \gamma J_T)$  implemented this way increases the circuit depth by $|T|-1$. In principle this can be reduced by replacing CNOT ladder circuits with the circuits introduced in \cite{baumer_measurement-based_2025,tserkisDepthOptimizationCNOT2026}. However, this comes at the expense of doubling the CNOT count and introducing ancilla qubits.}

To implement all terms of the HUBO-Hamiltonian $\sum_{T \subseteq \{1, \ldots, d\}} J_T \hat{Z}_T$, we construct a circuit that creates parity states $\ket{\oplus_{t \in T} b_t}$ and acts $R_Z$ gates on these states for all subsets $T \subseteq \{1, \ldots, d\}$ at least once. This can, in principle, be done by using circuits like in \figureLabel \ref{fig:exponentiated_cost} sequentially. However, this is suboptimal in the number of CNOT gates. The construction of parity states $\ket{\oplus_{t \in T} b_t}$ for all subsets $T \subseteq \{1, \ldots, d\}$ with minimal CNOT gate costs is a routing problem, and as shown in \cite{verchereOptimizingVariationalCircuits2023a} and \cite{welchEfficientQuantumCircuits2014} the Gray Code implementation is the optimal approach to it.

The key observation for optimizing the circuit is that once the parity state corresponding to a subset $T_1 \subseteq \{1,\ldots,d\}$ is prepared, the parity state for any subset $T_2 \subseteq \{1,\ldots,d\}$ differing from $T_1$ by exactly one element can be obtained by a single additional operation. In particular, the phase factor $\exp(i J_{T_2}\hat{Z}_{T_2})$ is implemented by applying a single CNOT gate, with the differing element as control and the \response{qubit} encoding $\bigoplus_{a\in T_1} b_a$ as target.

To introduce the Gray Code-based circuit, we denote each subset $T_1 \subseteq \{1, \ldots, d\}$ through a binary string $\mathbf{g}$ of length $d$, where the $j$-th element is $1$ if $j \in T_1$ and $0$ otherwise. If two subsets $T_1, T_2$ differ in exactly one element, the corresponding binary strings $\mathbf{g}_1, \mathbf{g}_2$ differ in exactly one bit.

Binary strings of length $d$ map naturally to $d$-dimensional hypercubes, and implementing all subsets of $\{1, \cdots, d\}$ is equivalent to traversing all vertices of this hypercube. Each vertex of the hypercube corresponds to one binary string and therefore to one subset $T \subseteq \{1, \ldots, d\}$. Two neighboring vertices differ by one element. The construction of parity states $\ket{\oplus_{i \in T} b_i}$ for all subsets $T \subseteq \{1, \ldots, d\}$ with minimal CNOT gate costs, now amounts to traversing the hypercube with as few steps as possible.

\begin{figure}
  \centering
  \includegraphics[width=0.5\columnwidth]{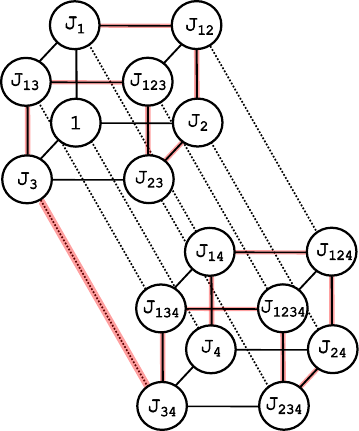}
  \caption{Gray code traversal for optimal implementation of a 4-qubit HUBO Hamiltonian. Each vertex of the 4-dimensional hypercube represents a subset $T \subseteq \{1,2,3,4\}$ of qubits, with subscripts indicating which qubits are included. The red path shows the Gray code sequence starting from the coefficient \response{$J_1$} (bitstring $1000$, the bitstring $0000$ corresponds to a global phase and is ignored). Each edge corresponds to flipping exactly one bit, enabling implementation of consecutive terms with only one additional CNOT gate per term. This systematic traversal visits all $2^4 - 1 = 15$ vertices exactly once, requiring only $2^4 - 2 = 14$ CNOT gates total for the complete HUBO Hamiltonian $\sum_{T \subseteq \{1,2,3,4\}} J_T \hat{Z}_T$.}
  \label{fig:gray_code}
\end{figure}

The Gray code provides a systematic way to traverse the vertices of this hypercube. Crucially, the Gray code visits each vertex of the hypercube exactly once, thereby avoiding any redundant operations or inefficiencies that would arise from revisiting previously implemented terms. Starting from the vertex corresponding to the global phase, i.e., the bitstring $\mathbf{0} = (0, 0, \ldots, 0)$ and following the Gray code sequence, we can implement all terms in the HUBO Hamiltonian using only one CNOT gate and one $R_Z$ gate per term. This is illustrated in \figureLabel \ref{fig:gray_code} for a four-qubit HUBO Hamiltonian. Welch et al.~\autocite{welchEfficientQuantumCircuits2014} have proven that this Gray code implementation is optimal in the sense that it requires the minimum number of CNOT gates to implement all terms in the HUBO Hamiltonian.

\begin{figure}
  \centering
  \includegraphics[width=100mm]{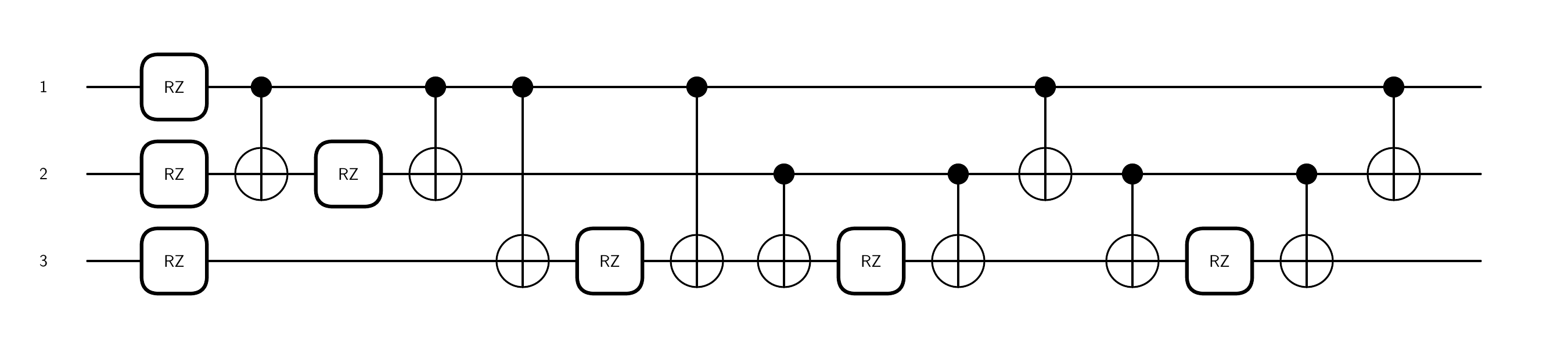}
  \caption{Non-optimized implementation of the cost unitary for a three-qubit HUBO Hamiltonian $\sum_{T \subseteq \{1,2,3\}} J_T \hat{Z}_T$, where $T$ runs over all subsets of the three qubits. Each term is implemented sequentially using parity circuits as shown in \figureLabel \ref{fig:exponentiated_cost}, resulting in $2^4 (4- 2) + 2 = 10$ CNOT gates and $2^4 - 1 = 7$ $R_Z$ gates. This method incurs a higher CNOT count than the Gray code-based optimal circuit presented in \figureLabel \ref{fig:optimal_circuit}, highlighting the benefit of circuit optimization for efficient quantum computation.}
  \label{fig:non_optimal_circuit}
\end{figure}

\begin{figure}
  \centering
  \includegraphics[width=35mm]{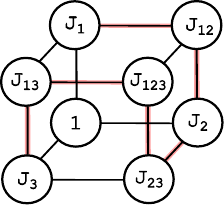}
  \includegraphics[width=100mm]{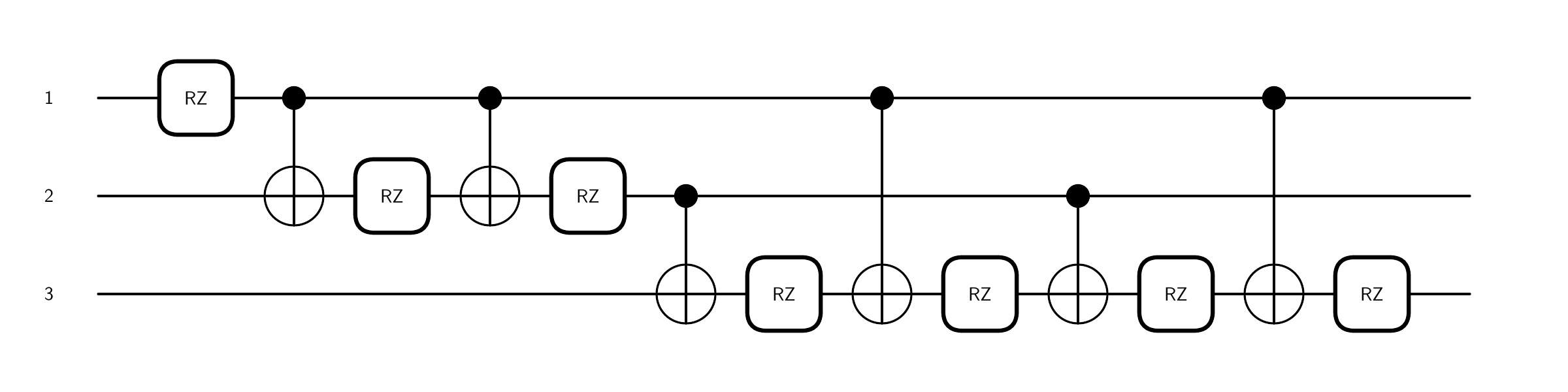}
  \caption{Optimal quantum circuit for implementing the cost unitary of a three-qubit HUBO Hamiltonian using the Gray code approach. Here, each subset $T \subseteq \{1,2,3\}$ defines a term $J_T \hat{Z}_T$ in the Hamiltonian.
    The upper panel illustrates the Gray code traversal of the three-dimensional hypercube, enabling efficient implementation with one additional CNOT and one $R_Z$ gate per term. For three qubits, this results in $2^3 - 2 = 6$ CNOT gates and $2^3 - 1 = 7$ $R_Z$ gates for all nontrivial subsets $T$, demonstrating the reduction in CNOT gates achieved by this optimized construction. Compared to the non-optimized implementation shown in \figureLabel \ref{fig:non_optimal_circuit}, this approach significantly reduces the CNOT gate count.}
  \label{fig:optimal_circuit}
\end{figure}

In practice, this means that a HUBO Hamiltonian of the form $\sum_{T \subseteq \{1, \ldots, d\}} J_T \hat{Z}_T$ can be implemented with $2^\abs{d} - 2$ CNOT gates and $2^\abs{d} - 1$ $R_Z$ gates. The \response{sequential} approach requires $2^\abs{d}(\abs{d}-2) + 2$ CNOT gates.
An illustration of the optimized circuit for the three-qubit HUBO Hamiltonian, in comparison to the non-optimized implementation shown in \figureLabel \ref{fig:non_optimal_circuit}, is presented in \figureLabel \ref{fig:optimal_circuit}.


    \section{Derivation: Scaling of Quantum Resources for QUBO and HUBO Formulations}
\label{sec:appendix_derivation_scaling}
In this section, we derive scaling relations for the quantum resources required to encode a quadratic COP in the QUBO and HUBO encodings.
The COP we consider has
a quadratic objective function of the form
\begin{equation}
    C(\solutionString) = \sum_{1 \leq i \leq n} c_1(i, s_i) + \sum_{1 \leq i < j \leq n} c_2(i, j, s_i, s_j).
    \label{eqn:classical_objective_function_appendix}
\end{equation}
We will consider the most general instance of this problem, where the linear coefficients $c_1(i, v)$ are nonzero for all variables $i$ and all values $v$, and the quadratic coefficients $c_2(i, j, v, w)$ are nonzero for all variable combinations $i \neq j$ and all value pairs $v, w$. This instance then has $n \cdot m$ linear coefficients and $\binom{n}{2} \cdot m^2$ quadratic coefficients.
Of course, in practice, the objective function often has a simpler form, but here we consider the worst-case scaling of the quantum resources required for the problem.

\subsection{QUBO formulation}
\label{sec:appendix_derivation_scaling_one_hot}
The QUBO encoding uses $n \cdot m$ qubits, where each qubit represents a unique variable-value pair.
The linear coefficients in the objective function give rise to one-body terms in the cost Hamiltonian. Since each one-body term is implemented using a single $R_Z$ gate, the number of $R_Z$ gates required to implement all linear terms is equal to the number of qubits. %

The quadratic coefficients translate to two-body terms in the cost Hamiltonian, each requiring a circuit with two CNOT gates and one $R_Z$ gate (see \figureLabel \ref{fig:exponentiated_cost}, left circuit).
Since there are $\binom{n}{2} \cdot m^2$ quadratic coefficients, the number of CNOT gates required to implement all quadratic terms is $2 \cdot \binom{n}{2} \cdot m^2$, and the number of $R_Z$ gates is $\binom{n}{2} \cdot m^2$.
The asymptotic scaling of the number of CNOT and $R_Z$ gates is therefore $\mathcal{O}(n^2 m^2)$.

\subsection{HUBO encoding}
In the HUBO encoding, each variable is represented by $d = \lceil \log_2(m) \rceil$ qubits.
Therefore, the total number of qubits required is $n \cdot \lceil \log_2(m) \rceil$. The linear coefficients in the objective function give rise to a Hamiltonian term $\sum_{T \subseteq T_i} J_{T} \hat{Z}_T$, where $T_i$ denotes the set of $d$ qubits encoding variable $i$.
This generates many-body interactions ranging from single-qubit terms (1st order) up to $d$-qubit terms ($d$th order) for each of the $n$ variables.

The quadratic coefficients for variable combinations $i \neq j$ produce a Hamiltonian term $\sum_{T_1 \subseteq T_i, T_2 \subseteq T_j} J_{T_1, T_2} \hat{Z}_{T_1 \cup T_2}$. Here, $T_1$ and $T_2$ are subsets of the qubit indices, $T_i$ and $T_j$, encoding variables $i$ and $j$, respectively.
Importantly, since $T_i \subseteq T_i \cup T_j$, the linear terms can be encoded simultaneously with the quadratic terms, eliminating the need for separate linear term implementation.
As discussed in Appendix~\ref{sec:appendix_cost_hamiltonian_circuit}, implementing the coefficients for all subsets of a set of qubit indices $T$ requires $2^{|T|} - 1$ $R_Z$ gates and $2^{|T|} - 2$ CNOT gates.
Since the union $T_i \cup T_j$ involves $2d$ qubits,
this translates to $2^{2d} - 1$ $R_Z$ gates and $2^{2d} - 2$ CNOT gates per variable pair.
With $\binom{n}{2}$ variable pairs, the total number of $R_Z$ gates required is $\binom{n}{2} \cdot (2^{2d} - 1)$ and the total number of CNOT gates is $\binom{n}{2} \cdot (2^{2d} - 2)$. With $d = \lceil \log_2(m) \rceil$, this gives us the asymptotic scaling of the CNOT and $R_Z$ gates as $\mathcal{O}(n^2 m^2)$.
\response{In contrast, the sequential implementation requires $2^{2d}(2d-1) + 2$ CNOT gates per variable pair. The CNOT gate count in the sequential implementation thus scales as $\mathcal{O}(n^2 m^2 \log(m))$.}

    \section{Technical Details on QAOA Experiments}
\label{sec:appendix_details_qaoa}
The exact computation of the approximation ratio in \equationLabel \eqref{eqn:approximation_ratio} requires summing over all feasible solutions, a process that scales exponentially with problem size in the considered benchmarks. Therefore, we estimate the approximation ratio using sampling-based methods. Specifically, the sum over all feasible states is replaced by a sum over sampled bitstrings,
\begin{equation}
    \sum_{\mathbf{b} \in B_\text{feas}} \rightarrow \sum_{\mathbf{b}\in B_\text{sample}} \delta(\mathbf{b}),
\end{equation}
where $\delta(\mathbf{b})$ equals one if the bitstring $\mathbf{b}$ satisfies all constraints, and zero otherwise. The required sample complexity is determined using Hoeffding’s inequality~\cite{hoeffdingProbabilityInequalitiesSums1963}. Employing $\abs{B_\text{sample}} = 10000$ samples in all experiments ensures a 99\% probability that the estimated value is within 2\% of the true approximation ratio.

Furthermore, for all benchmarked COP instances and QAOA depths, we performed 100 independent QAOA runs. This ensures that all reported averages and standard deviations are directly comparable across experiments. For the target approximation benchmarks, we ran the QAOA algorithm $100$ times and picked the QAOA experiment with the lowest layer requirements to reach the target threshold.

\subsection{Percentage Reduction Calculations}
\response{Throughout this work, percentage reductions in gate counts are computed as:
\begin{equation}
    \text{Reduction \%} = 100 \times \frac{\text{Gate count}_{\text{QUBO}} - \text{Gate count}_{\text{HUBO}}}{\text{Gate count}_{\text{QUBO}}},
\end{equation}
where a reduction of 100\% occurs when $\text{Gate count}_{\text{HUBO}} = 0$, as happens for single-variable instances when no CNOT gates are required. All reported percentage reductions refer to per-layer gate counts. Detailed per-layer gate counts (CNOT, $R_Z$, Hadamard, and $R_X$) for all benchmarked problem sizes and both encodings are provided in Table~\ref{tab:all_gate_counts_detailed}.}
\response{The Gray code circuit implementation is more efficient than the sequential implementation for a full HUBO Hamiltonian $\sum_{T \subseteq \{1, \ldots, d\}} J_T \hat{Z}_T$. However, when there are zero HUBO coefficients $J_T$ for some sets $T$ then in principle the sequential implementation can also be more CNOT.}
\response{For the MkCS instance with one vertex, we omit the HUBO scaling because, in the HUBO encoding, any sampled two-bit string is already optimal. Consequently, this instance can be solved with a zero-layer QAOA circuit.}

\begin{table}[h!]
    \centering
    \begin{tabular}{lcccccccc}
        \toprule
        Problem Instance  & \multicolumn{4}{c}{\textbf{QUBO}} & \multicolumn{4}{c}{\textbf{HUBO}}                                                      \\
        \cmidrule(lr){2-5} \cmidrule(lr){6-9}
                          & CNOT                              & $R_Z$                             & Hadamard & $R_X$ & CNOT & $R_Z$ & Hadamard & $R_X$ \\
        \midrule
        \multicolumn{9}{c}{\textit{Gate Assignment Problem (GAP, 4 gates fixed)}}                                                                      \\
        GAP (1 flight)    & 12                                & 10                                & 4        & 4     & 0    & 1     & 2        & 2     \\
        GAP (2 flights)   & 32                                & 24                                & 8        & 8     & 10   & 5     & 4        & 4     \\
        GAP (3 flights)   & 76                                & 50                                & 12       & 12    & 34   & 14    & 6        & 6     \\
        GAP (4 flights)   & 96                                & 64                                & 16       & 16    & 44   & 18    & 8        & 8     \\
        GAP (5 flights)   & 140                               & 90                                & 20       & 20    & 68   & 27    & 10       & 10    \\
        \midrule
        \multicolumn{9}{c}{\textit{Maximum $k$-Colorable Subgraph (MkCS, 4 colors fixed)}}                                                             \\
        MkCS (1 vertex)   & 12                                & 10                                & 4        & 4     &      &       &          &       \\
        MkCS (2 vertices) & 32                                & 24                                & 8        & 8     & 10   & 3     & 4        & 4     \\
        MkCS (3 vertices) & 52                                & 38                                & 12       & 12    & 20   & 6     & 6        & 6     \\
        MkCS (4 vertices) & 88                                & 60                                & 16       & 16    & 50   & 15    & 8        & 8     \\
        MkCS (5 vertices) & 132                               & 86                                & 20       & 20    & 90   & 27    & 10       & 10    \\
        \midrule
        \multicolumn{9}{c}{\textit{Integer Programming (IP, 4 values fixed)}}                                                                          \\
        IP (1 variable)   & 12                                & 10                                & 4        & 4     & 0    & 2     & 2        & 2     \\
        IP (2 variables)  & 42                                & 29                                & 8        & 8     & 8    & 7     & 4        & 4     \\
        IP (3 variables)  & 90                                & 57                                & 12       & 12    & 24   & 18    & 6        & 6     \\
        IP (4 variables)  & 120                               & 76                                & 16       & 16    & 38   & 31    & 8        & 8     \\
        IP (5 variables)  & 150                               & 95                                & 20       & 20    & 46   & 37    & 10       & 10    \\
        \bottomrule
    \end{tabular}
    \caption{Per-layer gate counts (CNOT, $R_Z$, Hadamard, $R_X$) for all benchmarked problem instances and sizes, comparing QUBO and HUBO encodings.}
    \label{tab:all_gate_counts_detailed}
\end{table}

\subsection{One-Hot Constraint Contribution in QUBO Encoding}\label{sec:oh_resource_requirements}
\response{To understand the source of gate-count reduction, we report in Table~\ref{tab:one_hot_constraint_gates} the per-layer CNOT and $R_Z$ gate counts required solely for implementing the one-hot penalty terms in the QUBO encoding. The one-hot penalty $\lambda(1 - \sum_{v=1}^{m} x_{i,v})^2$ for each variable expands to $m$ linear terms and $m(m-1)$ two-body terms. For all three problems with $n$ variables and values $m=4$, this yields $12 n$ CNOT gates and $10 n$ RZ gates per variable, independent of the problem type. Note that in the full QUBO encoding, some of these gates may be shared or merged with gates implementing the objective function and problem-specific constraints; therefore, the total gate count is not simply the sum of the one-hot and problem-specific contributions.}

\begin{table}[h!]
    \centering
    \begin{tabular}{lcc}
        \toprule
        Number of Variables & CNOT & $R_Z$ \\
        \midrule
        1                   & 12   & 10    \\
        2                   & 24   & 20    \\
        3                   & 36   & 30    \\
        4                   & 48   & 40    \\
        5                   & 60   & 50    \\
        \bottomrule
    \end{tabular}
    \caption{Per-layer CNOT and $R_Z$ gate counts for one-hot penalty terms in the QUBO encoding. These counts are identical for all three benchmarked problems (GAP, MkCS, and IP) since they all use $m=4$ values per variable.}
    \label{tab:one_hot_constraint_gates}
\end{table}

    \section{Details on the GAP Instance}
\label{sec:appendix_gap_benchmark_details}

\begin{table}
\centering
\begin{tabular}{lccccc}
\hline
              & Check-in/Luggage claim & Gate 1 & Gate 2 & Gate 3 & Gate 4 \\
\hline
Check-in/Luggage claim & 0 & 10 & 10 & 20 & 20 \\
Gate 1 & 10 & 0 & 20 & 20 & 20 \\
Gate 2 & 10 & 20 & 0 & 20 & 20 \\
Gate 3 & 20 & 20 & 20 & 0 & 1 \\
Gate 4 & 20 & 20 & 20 & 1 & 0 \\
\hline
\end{tabular}
\caption{Walking times (in minutes) between check-in/luggage claim and airport gates 1-4 for the benchmarked GAP instance.}
\label{tab:walking_times}
\end{table}
\begin{table}
\centering
\begin{tabular}{lc}
\hline
Aircraft i & $p^{arr}_i + p^{dep}_i$ \\
\hline
0 & 75 \\
1 & 74 \\
2 & 62 \\
3 & 88 \\
4 & 61 \\
\hline
\end{tabular}
\caption{Number of arriving and departing passengers for each of the five aircraft in the benchmarked GAP instance.}
\label{tab:ac_passengers}
\end{table}
For benchmarking, we use a GAP instance with $n=5$ flights and $m=4$ airport-gates, numbered 1 to 4, as shown in Fig.~\ref{fig:objective_values_gap}. The numbers of arriving and departing passengers per flight are given in Table~\ref{tab:ac_passengers}, with values ranging from 61 to 88. Out of the total, there are 13 transfer passengers between flights 1 and 4, and 29 transfer passengers between flights 0 and 2. Transfer passengers contribute to the quadratic cost term $c_2(i,j,v,w)$ in \equationLabel \eqref{eqn:gap_quadratic_term}, while the remaining passengers are included in the linear term $c_1(i,v)$ in \equationLabel \eqref{eqn:gap_linear_term}.

The walking times between check-in/luggage claim and the airport-gates, as well as in between the airport-gates, are given in Table~\ref{tab:walking_times}. In this instance, the walking times from each gate to check-in and from each gate to luggage claim are identical. For this reason, there is no difference in the cost contribution from arriving or departing passengers. Both groups affect the objective function in the same way, hence we don't distinguish between them in Table~\ref{tab:ac_passengers}.

Gate assignments are further constrained by overlapping flight schedules. Any flights that overlap cannot be assigned to the same gate. The overlapping flight pairs in this instance are $\{(0,1),\ (1,2),\ (2,3),\ (3,4)\}$.

\end{appendices}

\bibliography{sn-bibliography}%

\begin{thebibliography}{10}
\providecommand{\doi}[1]{\url{https://doi.org/#1}}
\bibcommenthead

\bibitem[\protect\citeauthoryear{Marzec}{2016}]{marzecPortfolioOptimizationApplications2016}
Marzec M.
\newblock Portfolio Optimization: Applications in Quantum Computing.
\newblock John Wiley \& Sons, Ltd; 2016.
\newblock Available from: \url{https://onlinelibrary.wiley.com/doi/abs/10.1002/9781118593486.ch4}.

\bibitem[\protect\citeauthoryear{Perdomo-Ortiz et~al.}{2012}]{perdomo-ortizFindingLowenergyConformations2012}
Perdomo-Ortiz A, Dickson N, Drew-Brook M, Rose G, Aspuru-Guzik A.
\newblock Finding Low-Energy Conformations of Lattice Protein Models by Quantum Annealing.
\newblock Scientific Reports. 2012 Aug;2(1):571.
\newblock \doi{10.1038/srep00571}.

\bibitem[\protect\citeauthoryear{Boucherie et~al.}{2021}]{doi:10.1142/12343}
Boucherie RJ, Braaksma A, Tijms H.
\newblock Operations Research.
\newblock WORLD SCIENTIFIC; 2021.

\bibitem[\protect\citeauthoryear{Crescenzi and Kann}{1995}]{crescenziCompendiumNPOptimization1995}
Crescenzi P, Kann V.
\newblock A Compendium of {{NP}} Optimization Problems.
\newblock Springer Berlin Heidelberg; 1995.
\newblock Available from: \url{https://cs.pwr.edu.pl/zielinski/lectures/om/compendium.pdf}.

\bibitem[\protect\citeauthoryear{Fu and Anderson}{1986}]{fuApplicationStatisticalMechanics1986}
Fu Y, Anderson PW.
\newblock Application of Statistical Mechanics to {{NP-complete}} Problems in Combinatorial Optimisation.
\newblock Journal of Physics A: Mathematical and General. 1986 Jun;19(9):1605.
\newblock \doi{10.1088/0305-4470/19/9/033}.

\bibitem[\protect\citeauthoryear{Kirkpatrick et~al.}{1983}]{kirkpatrickOptimizationSimulatedAnnealing}
Kirkpatrick S, Gelatt J C~D, Vecchi MP.
\newblock Optimization by Simulated Annealing.
\newblock Science. 1983;220(4598):671--680.
\newblock \doi{10.1126/science.220.4598.671}.

\bibitem[\protect\citeauthoryear{Glover et~al.}{1993}]{gloverUsersGuideTabu1993}
Glover F, TaiUard E, de~Werra D.
\newblock A User's Guide to Tabu Search.
\newblock Annals of Operations Research. 1993 Mar;41(1):1--28.
\newblock \doi{10.1007/BF02078647}.

\bibitem[\protect\citeauthoryear{{IBM}}{2024}]{IBMILOGCPLEX2024a}
{IBM}.
\newblock {IBM ILOG CPLEX Optimization Studio}; 2024.
\newblock Webpage.
\newblock Available from: \url{https://www.ibm.com/products/ilog-cplex-optimization-studio}.

\bibitem[\protect\citeauthoryear{{Gurobi}}{2025}]{LeaderDecisionIntelligence}
{Gurobi}.
\newblock Gurobi Optimization; 2025.
\newblock Webpage.
\newblock Available from: \url{https://www.gurobi.com/}.

\bibitem[\protect\citeauthoryear{Xu and Liberti}{2024}]{xuRelaxationsBinaryPolynomial2024}
Xu L, Liberti L.
\newblock Relaxations for binary polynomial optimization via signed certificates; 2024.
\newblock ArXiv preprint.
\newblock Available from: \url{https://arxiv.org/abs/2405.13447}.

\bibitem[\protect\citeauthoryear{Puchinger et~al.}{2010}]{puchingerMultidimensionalKnapsackProblem2010}
Puchinger J, Raidl GR, Pferschy U.
\newblock The {{Multidimensional Knapsack Problem}}: {{Structure}} and {{Algorithms}}.
\newblock INFORMS Journal on Computing. 2010 May;22(2):250--265.
\newblock \doi{10.1287/ijoc.1090.0344}.

\bibitem[\protect\citeauthoryear{Packebusch and Mertens}{2016}]{packebuschLowAutocorrelationBinary2016}
Packebusch T, Mertens S.
\newblock Low {{Autocorrelation Binary Sequences}}.
\newblock Journal of Physics A: Mathematical and Theoretical. 2016 Apr;49(16):165001.
\newblock \doi{10.1088/1751-8113/49/16/165001}.
\newblock {\href{https://arxiv.org/abs/1512.02475}{{arxiv:1512.02475}}}. {[cond-mat]}.

\bibitem[\protect\citeauthoryear{Danilova et~al.}{2022}]{danilovaRecentTheoreticalAdvances2022}
Danilova M, Dvurechensky P, Gasnikov A, Gorbunov E, Guminov S, Kamzolov D, et~al.
\newblock Recent {{Theoretical Advances}} in {{Non-Convex Optimization}}.
\newblock In: Nikeghbali A, Pardalos PM, Raigorodskii AM, Rassias MT, editors. High-{{Dimensional Optimization}} and {{Probability}}: {{With}} a {{View Towards Data Science}}. Cham: Springer International Publishing; 2022. p. 79--163.

\bibitem[\protect\citeauthoryear{Burer and Letchford}{2012}]{burerNonconvexMixedintegerNonlinear2012}
Burer S, Letchford AN.
\newblock Non-Convex Mixed-Integer Nonlinear Programming: {{A}} Survey.
\newblock Surveys in Operations Research and Management Science. 2012 Jul;17(2):97--106.
\newblock \doi{10.1016/j.sorms.2012.08.001}.

\bibitem[\protect\citeauthoryear{Floudas et~al.}{2004}]{floudasGlobalOptimization21st2005}
Floudas CA, Akrotiriankis IG, Caratzoulas S, Meyer CA, Kallrath J.
\newblock Global Optimization in the 21st Century: {{Advances}} and Challenges.
\newblock Computers \& Chemical Engineering. 2004 May;29(6):1185--1202.
\newblock \doi{10.1016/j.compchemeng.2005.02.006}.

\bibitem[\protect\citeauthoryear{Albash and Lidar}{2018}]{albashAdiabaticQuantumComputation2018a}
Albash T, Lidar DA.
\newblock Adiabatic Quantum Computation.
\newblock Reviews of Modern Physics. 2018 Jan;90(1):015002.
\newblock \doi{10.1103/RevModPhys.90.015002}.
\newblock {\href{https://arxiv.org/abs/1611.04471}{{arXiv:1611.04471}}}. {[quant-ph]}.

\bibitem[\protect\citeauthoryear{Ebadi et~al.}{2022}]{ebadiQuantumOptimizationMaximum2022}
Ebadi S, Keesling A, Cain M, Wang TT, Levine H, Bluvstein D, et~al.
\newblock Quantum {{Optimization}} of {{Maximum Independent Set}} Using {{Rydberg Atom Arrays}}.
\newblock Science. 2022 Jun;376(6598):1209--1215.
\newblock \doi{10.1126/science.abo6587}.
\newblock {\href{https://arxiv.org/abs/2202.09372}{{arXiv:2202.09372}}}. {[quant-ph]}.

\bibitem[\protect\citeauthoryear{Farhi et~al.}{2014}]{farhiQuantumApproximateOptimization2014}
Farhi E, Goldstone J, Gutmann S.
\newblock A Quantum Approximate Optimization Algorithm; 2014.
\newblock ArXiv preprint.
\newblock Available from: \url{https://arxiv.org/abs/1411.4028}.

\bibitem[\protect\citeauthoryear{Lucas}{2014}]{lucasIsingFormulationsMany2014}
Lucas A.
\newblock Ising Formulations of Many {{NP}} Problems.
\newblock Frontiers in Physics. 2014 Feb;2.
\newblock \doi{10.3389/fphy.2014.00005}.

\bibitem[\protect\citeauthoryear{Goswami et~al.}{2024}]{goswamiSolvingOptimizationProblems2024}
Goswami K, Mukherjee R, Ott H, Schmelcher P.
\newblock Solving Optimization Problems with Local Light Shift Encoding on {{Rydberg}} Quantum Annealers.
\newblock Physical Review Research. 2024 Apr;6(2):023031.
\newblock \doi{10.1103/PhysRevResearch.6.023031}.
\newblock {\href{https://arxiv.org/abs/2308.07798}{{arXiv:2308.07798}}}. {[quant-ph]}.

\bibitem[\protect\citeauthoryear{Lai et~al.}{2025}]{lai2024arbitraryqubooptimizationanalysis}
Lai CT, Blank C, Schmelcher P, Mukherjee R.
\newblock Towards arbitrary QUBO optimization: analysis of classical and quantum-activated feedforward neural networks.
\newblock Machine Learning: Science and Technology. 2025 jun;6(2):025049.
\newblock \doi{10.1088/2632-2153/addb97}.

\bibitem[\protect\citeauthoryear{Zaman et~al.}{2022}]{zaman2021pyqubo}
Zaman M, Tanahashi K, Tanaka S.
\newblock PyQUBO: Python Library for Mapping Combinatorial Optimization Problems to QUBO Form.
\newblock IEEE Transactions on Computers. 2022;71(4):838--850.
\newblock \doi{10.1109/TC.2021.3063618}.

\bibitem[\protect\citeauthoryear{Dominguez et~al.}{2023}]{Dominguez_2023}
Dominguez F, Unger J, Traube M, Mant B, Ertler C, Lechner W.
\newblock Encoding-independent optimization problem formulation for quantum computing.
\newblock Frontiers in Quantum Science and Technology. 2023 Sep;2.
\newblock \doi{10.3389/frqst.2023.1229471}.

\bibitem[\protect\citeauthoryear{Montañez-Barrera et~al.}{2024}]{montanez-barreraUnbalancedPenalizationNew2024}
Montañez-Barrera JA, Willsch D, Maldonado-Romo A, Michielsen K.
\newblock Unbalanced Penalization: A New Approach to Encode Inequality Constraints of Combinatorial Problems for Quantum Optimization Algorithms.
\newblock Quantum Science and Technology. 2024 Apr;9(2):025022.
\newblock \doi{10.1088/2058-9565/ad35e4}.

\bibitem[\protect\citeauthoryear{Romero et~al.}{2025}]{romeroBiasFieldDigitizedCounterdiabatic2024}
Romero SV, Visuri AM, Gomez~Cadavid A, Simen A, Solano E, Hegade NN.
\newblock Bias-field digitized counterdiabatic quantum algorithm for higher-order binary optimization.
\newblock Communications Physics. 2025 Aug;8(1).
\newblock \doi{10.1038/s42005-025-02270-3}.

\bibitem[\protect\citeauthoryear{Glos et~al.}{2022}]{glosSpaceefficientBinaryOptimization2022}
Glos A, Krawiec A, Zimborás Z.
\newblock Space-Efficient Binary Optimization for Variational Quantum Computing.
\newblock npj Quantum Information. 2022 Apr;8(1):39.
\newblock \doi{10.1038/s41534-022-00546-y}.

\bibitem[\protect\citeauthoryear{Romero et~al.}{2025}]{romeroProteinFoldingAlltoall2025}
Romero SV, Gomez~Cadavid A, Nikačević P, Solano E, Hegade NN, Lopez-Ruiz MA, et~al.
\newblock Protein folding with an all-to-all trapped-ion quantum computer; 2025.
\newblock Arxiv Preprint.
\newblock Available from: \url{https://arxiv.org/abs/2506.07866}.

\bibitem[\protect\citeauthoryear{Yahui et~al.}{2023}]{chai2023simulatingflightgateassignment}
Yahui C, Epifanovsky E, Jansen K, Kaushik A, Kühn S.
\newblock Simulating the flight gate assignment problem on a trapped ion quantum computer; 2023.
\newblock ArXiv preprint.
\newblock Available from: \url{https://arxiv.org/abs/2309.09686}.

\bibitem[\protect\citeauthoryear{Nagies et~al.}{2025}]{nagiesBoostingQuantumAnnealing2025}
Nagies S, Geier KT, Akram J, Bantounas D, Johanning M, Hauke P.
\newblock Boosting Quantum Annealing Performance through Direct Polynomial Unconstrained Binary Optimization.
\newblock Quantum Science and Technology. 2025 Oct;10(3):035008.
\newblock \doi{10.1088/2058-9565/adcae6}.
\newblock {\href{https://arxiv.org/abs/2412.04398}{{arXiv:2412.04398}}}. {[quant-ph]}.

\bibitem[\protect\citeauthoryear{Wintersperger et~al.}{2023}]{Wintersperger_2023}
Wintersperger K, Dommert F, Ehmer T, Hoursanov A, Klepsch J, Mauerer W, et~al.
\newblock Neutral atom quantum computing hardware: performance and end-user perspective.
\newblock EPJ Quantum Technology. 2023 Aug;10(1).
\newblock \doi{10.1140/epjqt/s40507-023-00190-1}.

\bibitem[\protect\citeauthoryear{Fauseweh}{2024}]{fausewehQuantumManybodySimulations2024}
Fauseweh B.
\newblock Quantum Many-Body Simulations on Digital Quantum Computers: {{State-of-the-art}} and Future Challenges.
\newblock Nature Communications. 2024 Mar;15(1):2123.
\newblock \doi{10.1038/s41467-024-46402-9}.

\bibitem[\protect\citeauthoryear{Koch}{2025}]{Koch_PyHUBO_2025}
Koch F.
\newblock {PyHUBO}; 2025.
\newblock Github Repository.
\newblock Available from: \url{https://github.com/frederikKoch/PyHUBO}.

\bibitem[\protect\citeauthoryear{Schrijver}{2003}]{schrijverCombinatorailOptimization}
Schrijver A.
\newblock Combinatorial Optimization: Polyhedra and Efficiency. vol.~B.
\newblock Journal of Computer and System Sciences - JCSS; 2003.
\newblock Available from: \url{https://link.springer.com/book/9783540443896}.

\bibitem[\protect\citeauthoryear{Crainic et~al.}{2024}]{crainicCombinatorialOptimizationApplications2024}
Crainic TG, Gendreau M, Frangioni A, editors.
\newblock Combinatorial {{Optimization}} and {{Applications}}: {{A Tribute}} to {{Bernard Gendron}}. vol. 358 of International {{Series}} in {{Operations Research}} \& {{Management Science}}.
\newblock Cham: Springer Nature Switzerland; 2024.

\bibitem[\protect\citeauthoryear{Salehi et~al.}{2022}]{salehi_unconstrained_2022}
Salehi O, Glos A, Miszczak JA.
\newblock Unconstrained binary models of the travelling salesman problem variants for quantum optimization.
\newblock Quantum Information Processing. 2022 Jan;21(2):67.
\newblock \doi{10.1007/s11128-021-03405-5}.

\bibitem[\protect\citeauthoryear{López-Ibáñez et~al.}{2013}]{lopez-ibanez_travelling_2013}
López-Ibáñez M, Blum C, Ohlmann JW, Thomas BW.
\newblock The travelling salesman problem with time windows: {Adapting} algorithms from travel-time to makespan optimization.
\newblock Applied Soft Computing. 2013 Sep;13(9):3806--3815.
\newblock \doi{10.1016/j.asoc.2013.05.009}.

\bibitem[\protect\citeauthoryear{Chen et~al.}{2025}]{chenSlackvariableApproachVariational2025}
Chen J, Westerheim H, Holmes Z, Luo I, Nuradha T, Patel D, et~al.
\newblock Slack-Variable Approach for Variational Quantum Semidefinite Programming.
\newblock Physical Review A. 2025 Aug;112(2):022607.
\newblock \doi{10.1103/lwxq-4myj}.

\bibitem[\protect\citeauthoryear{Glover et~al.}{2022}]{gloverQuantumBridgeAnalytics2022}
Glover F, Kochenberger G, Du Y.
\newblock Quantum Bridge Analytics {{I}}: A Tutorial on Formulating and Using {{QUBO}} Models.
\newblock Annals of Operations Research. 2022 Jul;314(1):141--183.
\newblock \doi{10.1007/s10479-022-04634-2}.

\bibitem[\protect\citeauthoryear{Bouras et~al.}{2014}]{bourasAirportGateAssignment2014}
Bouras A, Ghaleb MA, Suryahatmaja US, Salem AM.
\newblock The {{Airport Gate Assignment Problem}}: {{A Survey}}.
\newblock The Scientific World Journal. 2014;2014:1--27.
\newblock \doi{10.1155/2014/923859}.

\bibitem[\protect\citeauthoryear{Chai et~al.}{2023}]{chaiFindingOptimalFlight2023}
Chai Y, Funcke L, Hartung T, Jansen K, Kühn S, Stornati P, et~al.
\newblock Towards {{Finding}} an {{Optimal Flight Gate Assignment}} on a {{Digital Quantum Computer}}.
\newblock Physical Review Applied. 2023 Dec;20(6):064025.
\newblock \doi{10.1103/PhysRevApplied.20.064025}.
\newblock {\href{https://arxiv.org/abs/2302.11595}{{arxiv:2302.11595}}}. {[quant-ph]}.

\bibitem[\protect\citeauthoryear{Bentert et~al.}{2019}]{bentertInductive$k$independentGraphs2019}
Bentert M, van Bevern R, Niedermeier R.
\newblock Inductive \$k\$-Independent Graphs and \$c\$-Colorable Subgraphs in Scheduling: {{A}} Review.
\newblock Journal of Scheduling. 2019 Feb;22(1):3--20.
\newblock \doi{10.1007/s10951-018-0595-8}.
\newblock {\href{https://arxiv.org/abs/1712.06481}{{arXiv:1712.06481}}}. {[cs]}.

\bibitem[\protect\citeauthoryear{Halld\'{o}rsson et~al.}{2004}]{halldorssonSpectrumSharingGames2004}
Halld\'{o}rsson MM, Halpern JY, Li LE, Mirrokni VS.
\newblock On Spectrum Sharing Games.
\newblock In: Proceedings of the Twenty-Third Annual {{ACM}} Symposium on {{Principles}} of Distributed Computing. St. John's Newfoundland Canada: ACM; 2004. p. 107--114.

\bibitem[\protect\citeauthoryear{Hertz et~al.}{2016}]{hertzConstructiveAlgorithmsPartial2016}
Hertz A, Montagné R, Gagnon F.
\newblock Constructive Algorithms for the Partial Directed Weighted Improper Coloring Problem.
\newblock Journal of Graph Algorithms and Applications. 2016 Feb;20(2):159--188.
\newblock \doi{10.7155/jgaa.00389}.

\bibitem[\protect\citeauthoryear{Koster and Scheffel}{2006}]{RoutingNetworkDimensioning2006}
Koster AMCA, Scheffel M.
\newblock A {{Routing}} and {{Network Dimensioning Strategy}} to Reduce {{Wavelength Continuity Conflicts}} in {{All-Optical Networks}}.
\newblock 10032. Optimization Online; 2006.
\newblock Available from: \url{https://optimization-online.org/?p=10032}.

\bibitem[\protect\citeauthoryear{Liu et~al.}{2025}]{liuEfficientHybridVariational2025}
Liu D, Li J, Cheng X, Zhang S, Chang Y, Yan L.
\newblock Efficient hybrid variational quantum algorithm for solving graph coloring problem; 2025.
\newblock ArXiv preprint.
\newblock Available from: \url{https://arxiv.org/abs/2504.21335}.

\bibitem[\protect\citeauthoryear{Quintero et~al.}{2021}]{quinteroCharacterizationQUBOReformulations2021}
Quintero R, Bernal D, Terlaky T, Zuluaga LF.
\newblock Characterization of QUBO reformulations for the maximum $k$-colorable subgraph problem; 2021.
\newblock ArXiv preprint.
\newblock Available from: \url{https://arxiv.org/abs/2101.09462}.

\bibitem[\protect\citeauthoryear{Wang et~al.}{2020}]{wang$XY$mixersAnalyticalNumerical2020}
Wang Z, Rubin NC, Dominy JM, Rieffel EG.
\newblock \${{XY}}\$-Mixers: Analytical and Numerical Results for {{QAOA}}.
\newblock Physical Review A. 2020 Jan;101(1):012320.
\newblock \doi{10.1103/PhysRevA.101.012320}.
\newblock {\href{https://arxiv.org/abs/1904.09314}{{arxiv:1904.09314}}}. {[quant-ph]}.

\bibitem[\protect\citeauthoryear{Streif et~al.}{2021}]{streifQuantumAlgorithmsLocal2021}
Streif M, Leib M, Wudarski F, Rieffel E, Wang Z.
\newblock Quantum Algorithms with Local Particle Number Conservation: Noise Effects and Error Correction.
\newblock Physical Review A. 2021 Apr;103(4):042412.
\newblock \doi{10.1103/PhysRevA.103.042412}.
\newblock {\href{https://arxiv.org/abs/2011.06873}{{arXiv:2011.06873}}}. {[quant-ph]}.

\bibitem[\protect\citeauthoryear{Sotirov et~al.}{2021}]{sotirovMaximum$k$colorableSubgraph2021}
Sotirov R, Kuryatnikova O, Vera J.
\newblock The maximum $k$-colorable subgraph problem and related problems; 2021.
\newblock ArXiv preprint.
\newblock Available from: \url{https://arxiv.org/abs/2001.09644}.

\bibitem[\protect\citeauthoryear{Wolsey}{2020}]{Formulations2020}
Wolsey L.
\newblock Integer Programming.
\newblock In: Integer {{Programming}}. John Wiley \& Sons, Ltd; 2020. p. 1--23.

\bibitem[\protect\citeauthoryear{Yves and Laurence}{2006}]{ProductionPlanningMixed2006}
Yves P, Laurence AW.
\newblock Production {{Planning}} by {{Mixed Integer Programming}}.
\newblock Springer {{Series}} in {{Operations Research}} and {{Financial Engineering}}. Springer New York; 2006.

\bibitem[\protect\citeauthoryear{Magat{\~a}o et~al.}{2002}]{magataoMixedIntegerProgramming2002}
Magat{\~a}o L, Arruda LVR, Neves~Jr F.
\newblock A {{Mixed Integer Programming Approach}} for {{Scheduling Commodities}} in a {{Pipeline}}.
\newblock In: Grievink J, {van Schijndel} J, editors. Computer {{Aided Chemical Engineering}}. vol.~10 of European {{Symposium}} on {{Computer Aided Process Engineering-12}}. Elsevier; 2002. p. 715--720.

\bibitem[\protect\citeauthoryear{Goswami et~al.}{2025}]{goswamiQuditbasedScalableQuantum2025}
Goswami K, Schmelcher P, Mukherjee R.
\newblock Qudit-based scalable quantum algorithm for solving the integer programming problem; 2025.
\newblock ArXiv preprint.
\newblock Available from: \url{https://arxiv.org/abs/2508.13906}.

\bibitem[\protect\citeauthoryear{Svensson et~al.}{2023}]{svenssonHybridQuantumClassicalHeuristic2023a}
Svensson M, Andersson M, Grönkvist M, Vikstål P, Dubhashi D, Ferrini G, et~al.
\newblock Hybrid {{Quantum-Classical Heuristic}} to {{Solve Large-Scale Integer Linear Programs}}.
\newblock Physical Review Applied. 2023 Sep;20(3):034062.
\newblock \doi{10.1103/PhysRevApplied.20.034062}.

\bibitem[\protect\citeauthoryear{Sharma and Lau}{2025}]{sharmaCuttingSlackQuantum2025}
Sharma M, Lau HC.
\newblock Cutting Slack: Quantum Optimization with Slack-Free Methods for Combinatorial Benchmarks; 2025.
\newblock ArXiv preprint.
\newblock Available from: \url{https://arxiv.org/abs/2507.12159}.

\bibitem[\protect\citeauthoryear{Tanahashi et~al.}{2019}]{tanahashi2019application}
Tanahashi K, Takayanagi S, Motohashi T, Tanaka S.
\newblock Application of Ising Machines and a Software Development for Ising Machines.
\newblock Journal of the Physical Society of Japan. 2019;88(6):061010.
\newblock \doi{10.7566/JPSJ.88.061010}.

\bibitem[\protect\citeauthoryear{Hadfield}{2021}]{hadfieldRepresentationBooleanReal2021}
Hadfield S.
\newblock On the Representation of {{Boolean}} and Real Functions as {{Hamiltonians}} for Quantum Computing.
\newblock ACM Transactions on Quantum Computing. 2021 Dec;2(4):1--21.
\newblock \doi{10.1145/3478519}.
\newblock {\href{https://arxiv.org/abs/1804.09130}{{arxiv:1804.09130}}}. {[quant-ph]}.

\bibitem[\protect\citeauthoryear{Farhi et~al.}{2001}]{farhiQuantumAdiabaticEvolution2001}
Farhi E, Goldstone J, Gutmann S, Lapan J, Lundgren A, Preda D.
\newblock A {{Quantum Adiabatic Evolution Algorithm Applied}} to {{Random Instances}} of an {{NP-Complete Problem}}.
\newblock Science. 2001 Apr;292(5516):472--475.
\newblock \doi{10.1126/science.1057726}.
\newblock {\href{https://arxiv.org/abs/quant-ph/0104129}{{arXiv:quant-ph/0104129}}}.

\bibitem[\protect\citeauthoryear{McArdle et~al.}{2019}]{mcardleVariationalAnsatzbasedQuantum2019}
McArdle S, Jones T, Endo S, Li Y, Benjamin SC, Yuan X.
\newblock Variational Ansatz-Based Quantum Simulation of Imaginary Time Evolution.
\newblock npj Quantum Information. 2019 Sep;5(1):75.
\newblock \doi{10.1038/s41534-019-0187-2}.

\bibitem[\protect\citeauthoryear{Beach et~al.}{2019}]{beachMakingTrottersSprint2019a}
Beach MJS, Melko RG, Grover T, Hsieh TH.
\newblock Making Trotters Sprint: {{A}} Variational Imaginary Time Ansatz for Quantum Many-Body Systems.
\newblock Physical Review B. 2019 Sep;100(9):094434.
\newblock \doi{10.1103/PhysRevB.100.094434}.

\bibitem[\protect\citeauthoryear{Love}{2020}]{loveCoolingImaginaryTime2020}
Love PJ.
\newblock Cooling with Imaginary Time.
\newblock Nature Physics. 2020 Feb;16(2):130--131.
\newblock \doi{10.1038/s41567-019-0709-z}.

\bibitem[\protect\citeauthoryear{Peruzzo et~al.}{2014}]{peruzzoVariationalEigenvalueSolver2014}
Peruzzo A, McClean J, Shadbolt P, Yung MH, Zhou XQ, Love PJ, et~al.
\newblock A Variational Eigenvalue Solver on a Photonic Quantum Processor.
\newblock Nature Communications. 2014 Jul;5(1):4213.
\newblock \doi{10.1038/ncomms5213}.

\bibitem[\protect\citeauthoryear{Tilly et~al.}{2022}]{tillyVariationalQuantumEigensolver2022}
Tilly J, Chen H, Cao S, Picozzi D, Setia K, Li Y, et~al.
\newblock The {{Variational Quantum Eigensolver}}: A Review of Methods and Best Practices.
\newblock Physics Reports. 2022 Nov;986:1--128.
\newblock \doi{10.1016/j.physrep.2022.08.003}.
\newblock {\href{https://arxiv.org/abs/2111.05176}{{arXiv:2111.05176}}}. {[quant-ph]}.

\bibitem[\protect\citeauthoryear{Zhou et~al.}{2020}]{zhouQuantumApproximateOptimization2020}
Zhou L, Wang ST, Choi S, Pichler H, Lukin MD.
\newblock Quantum {{Approximate Optimization Algorithm}}: {{Performance}}, {{Mechanism}}, and {{Implementation}} on {{Near-Term Devices}}.
\newblock Physical Review X. 2020 Jun;10(2):021067.
\newblock \doi{10.1103/PhysRevX.10.021067}.

\bibitem[\protect\citeauthoryear{Basso et~al.}{2022}]{bassoPerformanceLimitationsQAOA2022}
Basso J, Gamarnik D, Mei S, Zhou L.
\newblock Performance and Limitations of the {{QAOA}} at Constant Levels on Large Sparse Hypergraphs and Spin Glass Models.
\newblock In: 2022 {{IEEE}} 63rd {{Annual Symposium}} on {{Foundations}} of {{Computer Science}} ({{FOCS}}); 2022. p. 335--343.

\bibitem[\protect\citeauthoryear{Blekos et~al.}{2024}]{blekosReviewQuantumApproximate2024}
Blekos K, Brand D, Ceschini A, Chou CH, Li RH, Pandya K, et~al.
\newblock A Review on {{Quantum Approximate Optimization Algorithm}} and Its Variants.
\newblock Physics Reports. 2024 Jun;1068:1--66.
\newblock \doi{10.1016/j.physrep.2024.03.002}.
\newblock {\href{https://arxiv.org/abs/2306.09198}{{arXiv:2306.09198}}}. {[quant-ph]}.

\bibitem[\protect\citeauthoryear{Golden et~al.}{2023}]{goldenNumericalEvidenceExponential2023}
Golden J, Bärtschi A, O’Malley D, Eidenbenz S.
\newblock Numerical {{Evidence}} for {{Exponential Speed-up}} of {{QAOA}} over {{Unstructured Search}} for {{Approximate Constrained Optimization}}.
\newblock In: 2023 {{IEEE International Conference}} on {{Quantum Computing}} and {{Engineering}} ({{QCE}}); 2023. p. 496--505.

\bibitem[\protect\citeauthoryear{Weidenfeller et~al.}{2022}]{weidenfellerScalingQuantumApproximate2022}
Weidenfeller J, Valor LC, Gacon J, Tornow C, Bello L, Woerner S, et~al.
\newblock Scaling of the Quantum Approximate Optimization Algorithm on Superconducting Qubit Based Hardware.
\newblock Quantum. 2022 Dec;6:870.
\newblock \doi{10.22331/q-2022-12-07-870}.
\newblock {\href{https://arxiv.org/abs/2202.03459}{{arXiv:2202.03459}}}. {[quant-ph]}.

\bibitem[\protect\citeauthoryear{Kurowski et~al.}{2023}]{kurowskiApplicationQuantumApproximate2023}
Kurowski K, Pecyna T, Slysz M, R\'{o}\.~{z}ycki R, Walig\'{o}ra G, W\k{e}glarz J.
\newblock Application of Quantum Approximate Optimization Algorithm to Job Shop Scheduling Problem.
\newblock European Journal of Operational Research. 2023 Oct;310(2):518--528.
\newblock \doi{10.1016/j.ejor.2023.03.013}.

\bibitem[\protect\citeauthoryear{Wang et~al.}{2023}]{wangQuantumAlternatingOperator}
Wang SS, Liu HL, Song YQ, Gao F, Qin SJ, Wen QY.
\newblock Quantum alternating operator ansatz for solving the minimum exact cover problem.
\newblock Physica A: Statistical Mechanics and its Applications. 2023 Sep;626:129089.
\newblock \doi{10.1016/j.physa.2023.129089}.

\bibitem[\protect\citeauthoryear{Sachdeva et~al.}{2024}]{sachdevaQuantumOptimizationUsing2024}
Sachdeva N, Hartnett GS, Maity S, Marsh S, Wang Y, Winick A, et~al.
\newblock Quantum optimization using a 127-qubit gate-model IBM quantum computer can outperform quantum annealers for nontrivial binary optimization problems; 2024.
\newblock ArXiv preprint.
\newblock Available from: \url{https://arxiv.org/abs/2406.01743}.

\bibitem[\protect\citeauthoryear{Welch et~al.}{2014}]{welchEfficientQuantumCircuits2014}
Welch J, Greenbaum D, Mostame S, Aspuru-Guzik A.
\newblock Efficient Quantum Circuits for Diagonal Unitaries without Ancillas.
\newblock New Journal of Physics. 2014 Mar;16(3):033040.
\newblock \doi{10.1088/1367-2630/16/3/033040}.

\bibitem[\protect\citeauthoryear{Blondel et~al.}{2022}]{jaxopt_implicit_diff}
Blondel M, Berthet Q, Cuturi M, Frostig R, Hoyer S, Llinares-López F, et~al.
\newblock Efficient and Modular Implicit Differentiation; 2022.
\newblock Arxiv Preprint.
\newblock Available from: \url{https://arxiv.org/abs/2105.15183}.

\bibitem[\protect\citeauthoryear{Bergholm et~al.}{2022}]{bergholmPennyLaneAutomaticDifferentiation2022}
Bergholm V, Izaac J, Schuld M, Gogolin C, Ahmed S, Ajith V, et~al.
\newblock PennyLane: Automatic differentiation of hybrid quantum-classical computations; 2022.
\newblock ArXiv preprint.
\newblock Available from: \url{https://arxiv.org/abs/1811.04968}.

\bibitem[\protect\citeauthoryear{Schulz et~al.}{2024}]{schulzGuidedQuantumWalk2024}
Schulz S, Willsch D, Michielsen K.
\newblock Guided Quantum Walk.
\newblock Physical Review Research. 2024 Mar;6(1):013312.
\newblock \doi{10.1103/PhysRevResearch.6.013312}.
\newblock {\href{https://arxiv.org/abs/2308.05418}{{arxiv:2308.05418}}}. {[quant-ph]}.

\bibitem[\protect\citeauthoryear{Verchère et~al.}{2023}]{verchereOptimizingVariationalCircuits2023a}
Verchère Z, Elloumi S, Simonetto A.
\newblock Optimizing Variational Circuits for Higher-Order Binary Optimization; 2023.
\newblock ArXiv preprint.
\newblock Available from: \url{https://arxiv.org/abs/2307.16756}.

\bibitem[\protect\citeauthoryear{Amy et~al.}{2018}]{amyCNOTcomplexityCNOTPHASECircuits2018}
Amy M, Azimzadeh P, Mosca M.
\newblock On the {{CNOT-complexity}} of {{CNOT-PHASE}} Circuits.
\newblock Quantum Science and Technology. 2018 Sep;4(1):015002.
\newblock \doi{10.1088/2058-9565/aad8ca}.
\newblock {\href{https://arxiv.org/abs/1712.01859}{{arxiv:1712.01859}}}. {[quant-ph]}.

\bibitem[\protect\citeauthoryear{Bäumer and Woerner}{2025}]{baumer_measurement-based_2025}
Bäumer E, Woerner S.
\newblock Measurement-based long-range entangling gates in constant depth.
\newblock Physical Review Research. 2025 May;7(2):023120.
\newblock \doi{10.1103/PhysRevResearch.7.023120}.

\bibitem[\protect\citeauthoryear{Tserkis et~al.}{2026}]{tserkisDepthOptimizationCNOT2026}
Tserkis S, Umer M, Angelakis DG.
\newblock Depth optimization of {CNOT} ladder circuits; 2026.
\newblock ArXiv:2511.13256 [quant-ph].
\newblock Available from: \url{http://arxiv.org/abs/2511.13256}.

\bibitem[\protect\citeauthoryear{Hoeffding}{1963}]{hoeffdingProbabilityInequalitiesSums1963}
Hoeffding W.
\newblock Probability {{Inequalities}} for {{Sums}} of {{Bounded Random Variables}}.
\newblock Journal of the American Statistical Association. 1963 Mar;58(301):13--30.
\newblock \doi{10.1080/01621459.1963.10500830}.

\end{thebibliography}

\end{document}